\documentclass[12pt]{article}
\pdfoutput=1

\usepackage{booktabs}
\usepackage{tabulary}
\usepackage{multirow}
\usepackage{rotating}
\usepackage{epsfig}
\usepackage{psfrag}
\usepackage{latexsym}
\usepackage{indentfirst}
\usepackage{fancyhdr}
\usepackage{dsfont}
\usepackage{adjustbox}
\usepackage{amsmath}
\usepackage{amssymb}
\usepackage{amsfonts}
\usepackage{mathrsfs}
\usepackage{amsthm}
\usepackage{pifont}
\usepackage{dsfont}
\usepackage{multirow}
\usepackage{array}
\usepackage{chngpage}
\usepackage{longtable}
\usepackage{cite}
\usepackage{bbold}
\usepackage{color}
\usepackage{braket}
\usepackage{colordvi}
\usepackage{fancybox}
\usepackage[footnotesize]{caption2}
\usepackage{graphicx}
\usepackage[center,footnotesize,hang]{subfigure}
\usepackage{bm}
\usepackage{bbm}
\usepackage{url}
\usepackage[table]{xcolor}
\usepackage[colorlinks, linkcolor=red, anchorcolor=black, citecolor=green]{hyperref}
\usepackage{multirow}
\usepackage{harpoon}
\usepackage{textcomp}  
\usepackage{enumitem}
\usepackage{mathrsfs}  

\usepackage{tikz}
\usepackage{diagbox}
\usepackage{enumitem}

\newcommand{\PreserveBackslash}[1]{\let\E^c=\\#1\let\\=\E^c}
\newcolumntype{C}[1]{>{\PreserveBackslash\centering}p{#1}}
\newcolumntype{R}[1]{>{\PreserveBackslash\raggedleft}p{#1}}
\newcolumntype{L}[1]{>{\PreserveBackslash\raggedright}p{#1}}
\allowdisplaybreaks  

\newcommand{\bq}{\begin{eqnarray}}
\newcommand{\nq}{\end{eqnarray}}

\newcommand{\ignore}[1]{}

\numberwithin{equation}{section}

\begin{document}
\title{
\begin{flushright}
\hfill\mbox{\small USTC-ICTS/PCFT-21-41} \\[5mm]
\begin{minipage}{0.2\linewidth}
\normalsize
\end{minipage}
\end{flushright}
{\Large \bf
Modular flavor symmetry and vector-valued modular forms
\\[2mm]}
\date{}

\author{
Xiang-Gan Liu\footnote{E-mail: {\tt
hepliuxg@mail.ustc.edu.cn}},  \
Gui-Jun~Ding\footnote{E-mail: {\tt
dinggj@ustc.edu.cn}}
\\*[20pt]
\centerline{
\begin{minipage}{\linewidth}
\begin{center}
$^1${\it \small
Interdisciplinary Center for Theoretical Study and  Department of Modern Physics,\\
University of Science and Technology of China, Hefei, Anhui 230026, China}\\[2mm]
$^2${\it \small Peng Huanwu Center for Fundamental Theory, Hefei, Anhui 230026, China}

\end{center}
\end{minipage}}
\\[10mm]}}

\maketitle
\thispagestyle{empty}

\begin{abstract}

We revisit the modular flavor symmetry from a more general perspective. The scalar modular forms of principal congruence subgroups are extended to the vector-valued modular forms, then we have more possible finite modular groups including $\Gamma_N$ and $\Gamma'_N$ as the flavor symmetry. The theory of vector-valued modular forms provide a method of differential equation to construct the modular multiplets, and it also reveals the simple structure of the modular invariant mass models. We review the theory of vector-valued modular forms and give general results for the lower dimensional vector-valued modular forms. The general finite modular groups are listed up to order 72. We apply the formalism to construct two new lepton mass models based on the finite modular groups $A_4\times Z_2$ and $GL(2,3)$.

\end{abstract}

\clearpage

{\hypersetup{linkcolor=black}
\tableofcontents
}

\section{Introduction}

The Standard Model (SM) of particle physics is an extremely successful theory. It describes the interactions between fundamental particles including quarks and leptons. The SM can explain almost all experimental results and precisely predict a wide variety of phenomena. However, it does not predict neither the masses nor the mixing parameters of the fundamental particles . The origin of flavor mixing parameters and  hierarchical fermion mass is of the greatest puzzle of SM, see Ref.~\cite{Xing:2020ijf} for a overview of the flavor structure of SM. The flavor symmetry acting on flavour space is one of the few guiding principles to address this problem.
The flavor group $G_f$ can be choosed as continuous Lie groups or discrete  non-abelian finite groups to associate the fermions between different family. In order to get a realistic mass spectrum and flavor mixing pattern, flavor group $G_f$ must be broken down to different subgroups in different flavor sectors, which relies on a carefully designed potential and often introduces extra symmetries and the so-called flavons. A latest review can be seen in~\cite{Feruglio:2019ybq}.

Recently modular invariance has been proposed from a bottom-up perspective as a promising framework for understanding the flavor puzzle~\cite{Feruglio:2017spp}, in which modular symmetry based on the infinite discrete group $SL(2,\mathbb{Z})$ plays the role of flavor symmetry, but in a nonlinear way. In addition, the traditional flavons are replaced by modulus field $\tau$, and the modular invariance restricts the Yukawa couplings to be the modular forms, which are holomorphic functions of $\tau$.
If the integer weight modular forms of level $N$ are used as the building block, the finite modular group is $\Gamma_N\equiv\bar{\Gamma}/\bar{\Gamma}(N)$ or its double covering group $\Gamma'_N\equiv\Gamma/\Gamma(N)$~\cite{Liu:2019khw}.
Modular flavor symmetry has been exploited to explain the hierarchical masses and mixing patterns in the lepton and quark sectors, and many models have been built by using the groups $\Gamma_2\cong S_3$~\cite{Kobayashi:2018vbk,Kobayashi:2018wkl}, $\Gamma_3\cong A_4$~\cite{Feruglio:2017spp,Criado:2018thu,Kobayashi:2018vbk,
Kobayashi:2018scp,Okada:2018yrn,Kobayashi:2018wkl,Okada:2019uoy,Ding:2019zxk,Kobayashi:2019xvz,Asaka:2019vev,Gui-JunDing:2019wap,Zhang:2019ngf,King:2020qaj,Ding:2020yen,Asaka:2020tmo,Okada:2020brs,Yao:2020qyy,Okada:2021qdf,Nomura:2021yjb}, $\Gamma_4\cong S_4$~\cite{Penedo:2018nmg,Novichkov:2018ovf,deMedeirosVarzielas:2019cyj,Kobayashi:2019mna,King:2019vhv,Criado:2019tzk,Wang:2019ovr,Gui-JunDing:2019wap,Wang:2020dbp,Qu:2021jdy}, $\Gamma_5\cong A_5$~\cite{Novichkov:2018nkm,Ding:2019xna,Criado:2019tzk}, $\Gamma_7\cong \mathrm{PSL}(2,\mathbb{Z}_7)$~\cite{Ding:2020msi}, $\Gamma'_3\cong T'$~\cite{Liu:2019khw,Lu:2019vgm}, $\Gamma'_4\cong S'_4$~\cite{Liu:2020akv,Novichkov:2020eep}, $\tilde{\Gamma}_{4}\cong \tilde{S}_4$~\cite{Liu:2020msy}, $\Gamma'_5\cong A'_5$~\cite{Wang:2020lxk,Yao:2020zml}, $\Gamma'_6\cong S_3\times T'$~\cite{Li:2021buv} and $\tilde{\Gamma}_5\cong A'_5\times Z_5$~\cite{Yao:2020zml}. The possible role of fractional weight modular forms has been explored, and the finite modular group should be extend to the metaplectic cover $\widetilde{\Gamma}_N\equiv \widetilde{\Gamma}/\Gamma(N)$ should be considered~\cite{Liu:2020msy}. The modular forms of weights $k/2$ with finite metaplectic modular group $\tilde{\Gamma}_{4}\cong \tilde{S}_4$~\cite{Liu:2020msy} and the weight $k/5$ modular forms with finite metaplectic modular group $\tilde{\Gamma}_5\cong A'_5\times Z_5$~\cite{Yao:2020zml} have been studied. Furthermore, the modular invariance framework is extended to allow for multiple moduli, and the classical modular forms are replaced by more extensive automorphic forms~\cite{Ding:2020zxw}. Moreover, as the first nontrivial example, the symplectic modular symmetry is presented~\cite{Ding:2020zxw}, which is based on Siegel modular group $Sp(2g,\mathbb{Z})$ and is a natural extension of modular group $SL(2,\mathbb{Z})$.
The generalized CP symmetry can be consistently combined with symplectic modular symmetries for both single modulus with $g=1$~\cite{Novichkov:2019sqv,Baur:2019kwi,Baur:2019iai} and multiple moduli with $g\geq 2$~\cite{Ding:2021iqp},  so that all the coupling coefficients become real under the canonical CP basis and real Clebsch-Gordan coefficients. There have also been attempts to embed modular symmetry into Grand Unified Theories~\cite{deAnda:2018ecu,Kobayashi:2019rzp,Du:2020ylx,Zhao:2021jxg,Chen:2021zty,Ding:2021zbg,Ding:2021eva,Charalampous:2021gmf}.
Some semi-quantitative characteristics and formalism of fermion mass matrices near fixed points of $\tau$ have also been developed~\cite{Okada:2020ukr,Feruglio:2021dte,Novichkov:2021evw}. The modular flavor symmetry has been explored from top-down perspective, and some finite modular groups can be naturally realized in top-down constructions based on string theory~\cite{Baur:2019kwi,Kobayashi:2020hoc,Ohki:2020bpo,Nilles:2020kgo,Kikuchi:2020frp,Nilles:2020tdp,Kikuchi:2020nxn,Ishiguro:2020tmo,Nilles:2021glx,Almumin:2021fbk,Kikuchi:2021yog,Baur:2021bly}.

In the modular flavor symmetry~\cite{Feruglio:2017spp}, it is assumed that the Yukawa couplings are modular forms of the principal congruence subgroup $\Gamma(N)$ and the matter fields transform in representations of the homogeneous (inhomogeneous) finite modular group $\Gamma'_N$ ($\Gamma_N$). The purpose of this paper is to revisit the modular invariance framework from a more general perspective. We extend the modular form of $\Gamma(N)$ to vector-valued modular form in irreducible representation of $SL(2,\mathbb{Z})$, and we focus on the representations with finite image. The matter fields are characterized by their modular weights and transformation properties under $SL(2,\mathbb{Z})$. Then the finite modular groups acting as flavor symmetry emerge as the images of the involved representations. As a result, we have more candidates for the finite modular groups other than $\Gamma'_N$ and $\Gamma_N$, which are generally
the quotients of $SL(2,\mathbb{Z})$ modulo its normal subgroups of finite index. Moreover, from the view of vector-valued modular form, the structure of the modular form space becomes much simpler and we can see there are only finite inequivalent modular invariant models with a given finite modular group.

This paper is organized as follows. In section~\ref{sec:VVMF}, the theory of vector-valued modular forms is reviewed. In section~\ref{sec:resultVVMF}, we give the general forms of the low dimensional vector-valued modular forms by solving the modular linear differential equation. We list the general finite modular groups up to order 72 in section~\ref{sec:framework}, and the modular flavor symmetry is extended by considering the vector-valued modular forms of $SL(2,\mathbb{Z})$ instead of the scalar modular forms of $\Gamma(N)$. In section~\ref{sec:example}, we present two example models for lepton masses and mixing based on the finite modular groups $A_4\times Z_2$ and $GL(2, 3)$. Finally, we draw the conclusions in section~\ref{sec:conclusion}. In Appendix~\ref{app:2-d irrep}, we list all two-dimensional irreducible representations of $SL(2,\mathbb{Z})$ with finite image. The vector-valued modular form modules of $A_4\times Z_2$ are presented in Appendix~\ref{app:VVMF-modules-A4xZ2}. We explain why there are finite possible lepton models based on a finite modular group in Appendix~\ref{app:finiteness}. The finite modular groups $GL(2,3)$, the tensor product decomposition and the Clebsch-Gordan Coefficients are given in Appendix~\ref{app:GL(2,3)_group}.

\section{\label{sec:VVMF}The theory of vector-valued modular forms}
The special linear group $SL(2,\mathbb{Z})\equiv\Gamma$ called the (full) modular group consists of matrices with integer entries and determinant 1:
\begin{equation}
SL(2,\mathbb{Z})=\left\{\begin{pmatrix}
a &~ b \\ c &~ d
\end{pmatrix}  \Big| ad- bc=1\,,\quad  a,b,c,d \in \mathbb{Z}\,\right\}\,.
\end{equation}
It is an infinite discrete group and can be generated by two generators $S$ and $T$ obeying the following relations
\begin{equation}
\label{eq:SL2Zrelation}
S^4=(ST)^3=1\,, \quad S^2 T=T S^2\,,
\end{equation}
with
\begin{equation}
S=\begin{pmatrix}
0 &~ 1 \\ -1 &~ 0
\end{pmatrix}\,, \qquad T=\begin{pmatrix}
1 &~ 1 \\ 0 &~ 1
\end{pmatrix}\,.
\end{equation}
Note that $S^2=-\mathbb{1}_2$, where $\mathbb{1}_2$ denotes the two-dimensional identity matrix. The modular group $\Gamma$ acts on the complex upper half-plane $\mathcal{H}=\{\tau\in\mathbb{C}~|~\Im \tau >0\}$ by linear fractional transformation
\begin{equation}
\gamma \tau \equiv \frac{a\tau+b}{c\tau+d}\,,\qquad \gamma=\begin{pmatrix}
a & b \\c & d
\end{pmatrix}\in \Gamma\,.
\end{equation}
It is easy to find that $\gamma$ and $-\gamma$ give the same action on $\tau$ and thus the faithful action of the linear fractional transformation is given by the projective special linear group $PSL(2,\mathbb{Z})$ which is the quotient group $PSL(2,\mathbb{Z})\cong SL(2,\mathbb{Z})/\{\mathbb{1}_2, -\mathbb{1}_2\}$. Obviously $S^2=-\mathbb{1}_2$ commutes with all element of $SL(2,\mathbb{Z})$, consequently $\{\mathbb{1}_2, -\mathbb{1}_2\}$ is exactly the center of $SL(2,\mathbb{Z})$ and by Schur's Lemma $\rho(S^2)$ is proportional to an identity matrix for any representation $\rho$ of $SL(2,\mathbb{Z})$. Furthermore we have $\rho(S^2)=\pm \mathbb{1}_{d}$ because of $S^4=1$, where $d$ denotes the dimension of the representation $\rho$. We call the representations with $\rho(S^2)=\mathbb{1}_{d}$ and $\rho(S^2)=-\mathbb{1}_{d}$ as even and odd representations respectively.
Any irreducible representation is either even or odd.

\subsection{Vector-valued modular forms and free-module theorem}

The theory of vector-valued modular form (VVMF) first appeared in the work of Selberg in 1960s~\cite{selberg1965estimation}, but it is not until the last two decades that the general theory of VVMF has been systematically developed thanks to the work of Mason, Gannon, Franc and et al~\cite{knopp2004vector,bantay2007vector,marks2010structure,gannon2014theory,franc2016hypergeometric,franc2018structure}. In physics, VVMFs appear as characters in rational conformal field theory~\cite{Eholzer:1994ta,Dong:1997ea}.

Let $\rho$ denote a $d$-dimensional representation of $SL(2,\mathbb{Z})$.
A vector-valued modular form $Y(\tau)=(Y_1(\tau),\dots,Y_d(\tau))^T$ of weight $k$ in representation $\rho$ is a holomorphic vector-valued function in the upper
half-plane $\mathcal{H}$, and transforms as\footnote{In addition, $Y(\tau)$ is generally required to satisfy the so-called moderate growth condition, i.e., there is an integer $N$ such that for each components $Y_j(\tau)$ we have $|Y_j(\tau)|< \Im(\tau)^N $ for any fixed $\tau\in\mathcal{H}$ and $\Im(\tau)\gg 0$. In other word, $Y_j(\tau)$ is bounded as $\Im(\tau)\rightarrow+\infty$.}
\begin{equation}
\label{eq:defVVMF}
Y(\gamma\tau)=(c\tau+d)^k\rho(\gamma)Y(\tau)\,,
\end{equation}
where $k$ is called the modular weight and $\rho(\gamma)$ is the irreducible representation matrix of $SL(2,\mathbb{Z})$. In the present work, we will focus on the case that $k$ is non-negative integer and $\rho$ has finite image. The benefits are as follows:
\begin{itemize}

\item[~~\textbf{(\lowercase\expandafter{\romannumeral1})}]
{$\rho$ can be taken to be unitary since the image of $\rho$ is finite, and it is also completely reducible by Maschke theorem~\cite{rao2006linear}, i.e., it can be decomposed into a direct sum of irreducible representations. Therefore, we can further restrict $\rho$ to the unitary irreducible representations without loss of generality.}

\item[~~\textbf{(\lowercase\expandafter{\romannumeral2})}]{The order of $\rho(\gamma)$ is finite for any $\gamma\in SL(2,\mathbb{Z})$ such that we can always work in the basis where $\rho(T)$ is diagonal, and the eigenvalues of $\rho(T)$ are roots of unity.}

\item[~~\textbf{(\lowercase\expandafter{\romannumeral3})}]{Finiteness of the image of $\rho$ implies that the kernel $\ker(\rho)$ is the  finite index normal subgroup of $SL(2,\mathbb{Z})$.}
\end{itemize}
Notice that the automorphy factor $j^k(\gamma,\tau)\equiv(c\tau+d)^k$ is a monodrome function for integral $k$. If the modular weight $k$ is a rational number, the automorphy factor would be a multi-valued function and $\rho$ is a projective representation. Then $SL(2,\mathbb{Z})$ should be extended to its metaplectic cover group $Mp(2,\mathbb{Z})$ in order to avoid the multi-valued ambiguity~\cite{Liu:2020msy}. By taking $\gamma=S^2$ in Eq.~\eqref{eq:defVVMF} and noticing $S^2\tau=\tau$, we have
\begin{equation}
Y(\tau)=(-1)^k\rho(S^2)Y(\tau)\,,
\end{equation}
which leads to
\begin{equation}
\rho(S^2)=(-1)^k\mathbb{1}_{d}\,.
\end{equation}
Therefore the odd/even weight VVMFs are in odd/even irreps. It is remarkable that each component of VVMF $Y(\tau)$ in the representation $\rho$ is scalar modular form of $\ker(\rho)$, this is obvious when $\gamma$ in Eq.~\eqref{eq:defVVMF} is limited to $\ker(\rho)$ rather than $SL(2,\mathbb{Z})$. The converse is also true, i.e., scalar modular forms for normal subgroups of $SL(2,\mathbb{Z})$ can be organized into the VVMFs for $SL(2,\mathbb{Z})$, as shown in Ref.~\cite{Liu:2019khw}. Notice that VVMFs in representation $\rho$ are referred to as modular multiplets of the finite modular group $\Gamma/\ker(\rho)$ in the modular flavor symmetry~\cite{Feruglio:2017spp}.

As mentioned above, it is convenient to work in the $T$ diagonal basis and the representation matrix $\rho(T)$ can be written as
\begin{equation}
\rho(T)=\texttt{diag}(e^{2\pi i r_1},\dots,e^{2\pi i r_d})\,,\qquad 0\leq r_i < 1\,,
\end{equation}
where $r_i$ are rational numbers. In this $T$-diagonal basis, the moderate growth condition implies that every nonzero VVMF $Y(\tau)\in \mathcal{M}_k(\rho)$ has the following holomorphic $q$-expansion~\cite{marks2012fourier}
\begin{equation}
\label{eq:q-form}
Y(\tau)=\begin{pmatrix}
Y_1(\tau) \\ \vdots \\ Y_d(\tau)
\end{pmatrix}=\begin{pmatrix}
q^{r_1}\sum_{n\geq 0}a_1(n)q^n \\ \vdots \\ q^{r_d}\sum_{n\geq 0}a_d(n)q^n
\end{pmatrix},~~~~q=e^{2\pi i\tau} \,,
\end{equation}
where the coefficients $a_j(n)$ are generally complex.

As the simplest examples, it is easy to observe that VVMFs are reduced to the (scalar) modular forms of $SL(2,\mathbb{Z})$ for trivial representation $\rho=\mathbf{1}$. As a well-known result, the modular forms space $\mathcal{M}(\mathbf{1})=\bigoplus^{\infty}_{k=0}\mathcal{M}_k(\mathbf{1})$ is just a graded ring generated by the Eisenstein series $E_4(\tau)$ and $E_6(\tau)$~\cite{cohen2017modular}:
\begin{equation}
\label{eq:E4-E6} E_4(\tau)=1+240\sum_{n=1}^{\infty}\sigma_3(n)q^n \,,~~~E_6(\tau)=1-504\sum_{n=1}^{\infty}\sigma_5(n)q^n \,,
\end{equation}
where $\sigma_k(n) = \sum_{d|n} d^k$ is the sum of the $k$th power of the divisors of $n$. We will denote by $\mathcal{M}_k(\rho)$ the linear space of holomorphic VVMFs of weight $k$ in irrep $\rho$, and the direct sum $\mathcal{M}(\rho)=\bigoplus_{k=0}^{\infty}\mathcal{M}_k(\rho)$ stands for
all the  VVMFs in the irrep $\rho$. Obviuosly multiplying a VVMF $Y(\tau)\in \mathcal{M}_k(\rho)$ with a scalar modular form  $f(\tau)\in \mathcal{M}_{2m}(\mathbf{1})$ results in a new VVMF $f(\tau)Y(\tau)\in \mathcal{M}_{k+2m}(\rho)$. Hence the direct sums $\mathcal{M}(\rho) = \bigoplus^{\infty}_{k=0}\mathcal{M}_k(\rho)$ is a graded module over the ring $\mathcal{M}(\mathbf{1})=\mathbb{C}[E_4,E_6]$~\footnote{A module is a generalization of the notion of vector space, where scalars are on a ring and no longer has to be on a field. If a module $\mathcal{M}$ can be generated by a linearly independent set $X$, we say $\mathcal{M}$ is a free module and $X$ is a basis of $\mathcal{M}$. The cardinality of the basis $X$ is called the rank of the free module $\mathcal{M}$~\cite{dummit1991abstract}.}. It can be proved that $\mathcal{M}(\rho)$ is a free module whose rank is equal to the dimension of the representation $\rho$~\cite{marks2010structure}. This is the famous free-module theorem about VVMFs, and it is important to explicitly construct VVMFs, as shown in the following section.

\subsection{Modular linear differential equation}

Let us consider the modular differential operators acting on the VVMFs of weight $k$~\cite{Bruinier2008The},
\begin{equation}
D_k\equiv \theta - \frac{kE_2}{12}\,,\quad \theta\equiv q\frac{d}{dq}=\frac{1}{2\pi i} \frac{d}{d\tau},\quad k\in\mathbb{Z}\,,
\end{equation}
where $E_2(\tau)$ is the well-known quasi-modular Eisenstein series~\cite{cohen2017modular}
\begin{equation}
\label{eq:E2}
E_2(\tau)=1-24\sum_{n=1}^{\infty}\sigma_1(n)q^n \,.
\end{equation}
Notice that $E_2(\tau)$ is not a modular form of weight $2$, and it fulfills the following transformation formula
\begin{equation}
E_2(\gamma\tau)=(c\tau+d)^2 E_2(\tau) -\frac{6i}{\pi} c(c\tau+d)\,,
\end{equation}
where the last term violates the modularity. It is not hard to verify, $D_k$ preserves modularity of modular forms and increasing its weight by 2, i.e.
\begin{equation}
(D_k f)(\gamma\tau)=(c\tau+d)^{k+2}(D_k f)(\tau)\, \qquad\text{for} \quad f\in \mathcal{M}_k(\mathbf{1})\,,
\end{equation}
and
\begin{equation}
(D_k Y)(\gamma\tau)=(c\tau+d)^{k+2}\rho(\gamma)(D_k Y)(\tau)\, \qquad\text{for} \quad Y\in \mathcal{M}_k(\rho)\,.
\end{equation}
A very common and important example first observed by Ramanujan is~\cite{Bruinier2008The}
\begin{equation}
\label{eq:Ramanujan}
D_2(E_2)=-\frac{E_4}{12}\,,\qquad D_4(E_4)=-\frac{E_6}{3}\,,\quad D_6(E_6)=-\frac{E_4^2}{2}\,.
\end{equation}
For any two scalar modular forms $f$ and $g$ which are of weight $k$ and $m$ respectively, the modular derivative satisfies the Leibniz law as follows,
\begin{equation}
\label{eq:Leibniz law}
D_{k+m}(fg)=(D_k f)g +f(D_m g)\,.
\end{equation}
For $n \geq 0$ , the $n$th iteration of $D_k$ is denoted by
\begin{equation}
D^{n}_k\equiv D_{k+2(n-1)}\circ D_{k+2(n-2)}\circ\dots \circ D_k\,,
\end{equation}
For a generic $d-$dimensional VVMF $F=\left(f_1,\ldots, f_d\right)^{T}\in \mathcal{M}_k(\rho)$, then the $d+1$ VVMFs $\{F,D_k F,\dots,D_k^d F\}$ are linearly dependent over $\mathcal{M}(\mathbf{1})$ since the rank of the module $\mathcal{M}(\rho)$ is equal to $d$. Hence there must exist $M_j\in \mathcal{M}_{j}(\mathbf{1})$ such that the following equation on the module $\mathcal{M}(\rho)$ is satisfied
\begin{equation}
\label{eq:MLDE}
(D^d_k+M_2D_k^{d-1}+M_4D_k^{d-2}+\dots+M_{2(d-1)}D_k+M_{2d})f_i=0\,,
\end{equation}
The above equation is actually the so-called $d$th order modular linear differential equation (MLDE) in weight $k$. From the basic theory of linear differential equations, we know that a homogeneous linear differential equation of order $d$ has $d$ linearly independent solutions. Hence the solution system forms $d$ linearly independent components of $F$.

On the other hand, the set VVMFs $\{F,D_k F,\dots,D_k^{d-1} F\}$ are linearly independent over $\mathcal{M}(\mathbf{1})$, otherwise they would satisfy a MLDE of order at most $d-1$, for which each of the $d$ linearly independent components of $F$ is a solution, this contradicts the irreducibility of $F$. It is worth noting that linear independence of $\{F,D_k F,\dots,D_k^{d-1} F\}$ does not mean they can be used as a basis of module $\mathcal{M}(\rho)$. In fact, they generally span rank $d$ free submodule of $\mathcal{M}(\rho)$.
Unless the module $\mathcal{M}(\rho)$ is cyclic\footnote{A module $\mathcal{M}(\rho)$ is called cyclic if and only if there is $F$ of minimal weight $k_0$ such that $\{F,D_{k_0}F,\dots,D^{d-1}_{k_0}F\}$ generates $\mathcal{M}(\rho)$. We simply call $\rho$ is cyclic type in this case.} and $F$ is the VVMF of minimal weight. Interestingly, the module $\mathcal{M}(\rho)$ of irrep $\rho$ with $\dim\rho\leq 3$ is always cyclic, while there is only non-cyclic module $\mathcal{M}(\rho)$ for $\dim\rho>6$~\cite{franc2018structure}.

\begin{table}[t!]
\centering
\begin{tabular}{|c|c|c|c|c}\hline\hline
 Dimension of $\rho$ & Weight profile & Minimal weight $k_0$ \\ \hline
  1  & $(k_0)$ & $12\text{Tr}(L)$ \\ \hline
  2  & $(k_0,k_0+2)$ & $6\text{Tr}(L)-1$ \\ \hline	
  3  & $(k_0,k_0+2,k_0+4)$ & $4\text{Tr}(L)-2$ \\ \hline	
\multirow{2}{*}{4} & $(k_0,k_0+2,k_0+4,k_0+6)$ & $3\text{Tr}(L)-3$  \\ \cline{2-3}
		& $(k_0,k_0+2,k_0+2,k_0+4)$ & $3\text{Tr}(L)-2$  \\ \hline
\multirow{5}{*}{5} & $(k_0,k_0+2,k_0+4,k_0+6,k_0+8)$ & $(12\text{Tr}(L)-20)/5$  \\ \cline{2-3}
& $(k_0,k_0+2,k_0+4,k_0+4,k_0+6)$ & $(12\text{Tr}(L)-16)/5$  \\ \cline{2-3}	
& $(k_0,k_0+2,k_0+2,k_0+4,k_0+4)$ & $(12\text{Tr}(L)-12)/5$  \\ \cline{2-3}	
& $(k_0,k_0,k_0+2,k_0+2,k_0+4)$ & $(12\text{Tr}(L)-8)/5$  \\ \cline{2-3}
& $(k_0,k_0+2,k_0+2,k_0+4,k_0+6)$ & $(12\text{Tr}(L)-14)/5$  \\ \hline\hline
\end{tabular}
\caption{\label{tab:IrrepProfile} The weight profile and minimal weight of the module $\mathcal{M}(\rho)$ with $\dim\rho < 6$, and the parameter $L$ is defined by $\rho(T)\equiv e^{2\pi iL}$ which implies $L=\text{diag}(r_1, r_2,\ldots, r_d)$. There are two types of irreps for dimension $4$, namely cyclic and non-cyclic, and corresponding to $\rho(S^2)=-(-1)^{3\text{Tr}(L)\,\text{mod}\, 2}$ and $\rho(S^2)=(-1)^{3\text{Tr}(L)\,\text{mod}\, 2}$ respectively~\cite{franc2020constructions}. There are five types of irreps for dimension $5$, corresponding to $\text{Tr}(L)=\frac{5n+i}{12}$ with $i=0,1,2,3,4$ and $n\in\mathbb{Z}$.}
\end{table}

The structure of the module $\mathcal{M}(\rho)$ is characterized by the weights of a basis of $\mathcal{M}(\rho)$, they are denoted as a tuple $(k_0,\dots,k_{d-1})$ which is called the weight profile of $\rho$. Here $k_0$ is the minimal weight and it is the minimal integer such that $\mathcal{M}_{k_0}(\rho)\neq 0$. The minimal weights and the weight profiles of the module $\mathcal{M}(\rho)$ for $\dim\rho<6$ are summarized in the table~\ref{tab:IrrepProfile}, which is adopted from Ref.~\cite{franc2018structure}. With these weight profiles, we can get a convenient dimension formula by using the Hilbert-Poincar\'e series which is the formal power series for module $\mathcal{M}(\rho)$~\cite{marks2009classification}:
\begin{equation}
P(\mathcal{M}(\rho))(t)=\sum_{k\geq k_0} t^k \dim(\mathcal{M}_k(\rho))\,,
\end{equation}
since $\mathcal{M}(\rho)$ is a free module of rank $d$ over $\mathbb{C}[E_4,E_6]$, its Hilbert-Poincar\'e series has the following general form:
\begin{equation}
\label{eq:HilbertSeries}
P(\mathcal{M}(\rho))(t)=\frac{t^{k_0}+\dots+t^{k_{d-1}}}{(1-t^4)(1-t^6)}\,,
\end{equation}
where $(k_0,\dots,k_{d-1})$ is the weight profile of $\rho$.
From table~\ref{tab:IrrepProfile}, we can easily write the generating function~\eqref{eq:HilbertSeries} of each module $\mathcal{M}(\rho)$, and then read the corresponding dimension of $\mathcal{M}_k(\rho)$ from the power series expansion\footnote{The more general dimension formula of the linear space $\mathcal{M}_k(\rho)$ for $k\geq2$~\cite{candelori2017vector}
\begin{equation}
\label{eq:dim-formula}
\dim \mathcal{M}_k(\rho)=\frac{(5+k)\dim\rho}{12}+\frac{i^k\text{Tr}(\rho(S^3))}{4}+\frac{(-\omega^2)^k\text{Tr}(\rho(S^3T))}{3(1-\omega)}+\frac{\omega^k\text{Tr}(\rho((ST)^2))}{3(1-\omega^2)}-\text{Tr}(L)
\end{equation}
with $\omega=e^{2\pi i/3}$. There is still no general dimension formula for $k=1$, although Eq.~\eqref{eq:dim-formula} is still applicable in some cases.}.
Taking the three-dimensional case as an example, the Hilbert-Poincar\'e series of $\mathcal{M}(\rho)$ is
\begin{equation}
\frac{t^{k_0}(1+t^2+t^4)}{(1-t^4)(1-t^6)}=t^{k_0}+t^{k_0+2}+2t^{k_0+4}+2t^{k_0+6}+\dots\,,
\end{equation}
with $k_0=4\text{Tr}(L)-2$. Therefore we know that there is only one modular multiplet at the weights $k_0$ and $k_0+2$, two modular multiplets at the weights $k_0+4$ and $k_0+6$, and so on.

For these low-dimensional VVMFs, the free base of module $\mathcal{M}(\rho)$ can be inferred from the weight profile and the free module theorem. For instance, at lower dimensions $d=1,2,3,4$~\cite{marks2009classification}, we have
\begin{align}
\nonumber&\dim\rho=1: ~~\mathcal{M}(\rho)=\langle F \rangle  \,,\\
\nonumber&\dim\rho=2: ~~\mathcal{M}(\rho)=\langle F, ~D_{k_0}F \rangle \,,\\
\nonumber&\dim\rho=3: ~~\mathcal{M}(\rho)=\langle F, ~D_{k_0}F, ~D^2_{k_0}F \rangle \,,\\
&\dim\rho=4: ~~\mathcal{M}(\rho)=\begin{cases}
\langle F,~D_{k_0}F,~D^2_{k_0}F,~D^3_{k_0}F \rangle ~~\text{for cyclic type}~~
\rho \\
\langle F,~D_{k_0}F,~G,~D^2_{k_0}F \rangle ~~~~~~\,\text{for non-cyclic type}~~\rho
\end{cases}\,,
\end{align}
where $F$ in each case is the corresponding minimal weight VVMF, and $G$ denotes the another independent VVMF at weight $k_0+2$ in addition to $D_{k_0}F$.
With these bases, we can write any high-weight VVMF as a linear combination of them over the module $\mathcal{M}(\mathbf{1})$.
For $\dim\rho=3$ as an example, the VVMF $D^3_{k_0}F$ of weight $k_0+6$ can be written as
\begin{equation}
\label{eq:example3dMLDE}
D^3_{k_0} F = a E_4 D_{k_0}F + b E_6 F\,,
\end{equation}
where $a$ and $b$ is the free complex numbers.
This happens to be a $3$rd order MLDE as shown in Eq.~\eqref{eq:MLDE}, and it can be used to solve the concrete form of $F$, as demonstrated in section~\ref{sec:resultVVMF}.

\section{\label{sec:resultVVMF}Some results on low dimensional VVMFs}

In this section, we report the explicit form of one-dimensional, two-dimensional and three-dimensional VVMFs by solving the MLDE in Eq.~\eqref{eq:MLDE}.  It is worth mentioning that the MLDE of Eq.~\eqref{eq:MLDE} satisfied by 2-d and 3-d VVMF can be transformed into the generalized hypergeometric equation of the following form~\cite{franc2016hypergeometric}:
\begin{equation}
\label{eq:generalHyperDE}
\left[(\theta_K+\beta_1-1)\cdots (\theta_K+\beta_n-1) - K(\theta_K+\alpha_1)\cdots(\theta_K+\alpha_n)\right]f_i=0\,.
\end{equation}
with
\begin{equation}
\theta_K=K\frac{d}{dK}\,,\quad K(\tau)=1728/j(\tau)\,.
\end{equation}
Here $j(\tau)$ is the modular $j$-invariant~\cite{cohen2017modular} and it is a modular function of weight 0,
\begin{equation}
\label{eq:j-Delta}j(\tau)=\frac{E^3_4(\tau)}{\Delta(\tau)}\,,\quad \Delta(\tau)=\frac{E_4^3(\tau)-E_6^2(\tau)}{1728}\,,
\end{equation}
The function $j(\tau)$ has a simple pole at $i\infty$, and its $q$-expansion is
\begin{equation}
j(\tau)=q^{-1}+744+196884q+21493760q^2+\dots\,.
\end{equation}
When the numbers $\beta_{1},\dots,\beta_{n}$ are distinct mod $\mathbb{Z}$, the equation~\eqref{eq:generalHyperDE} has $n$ independent solutions which are given by the generalized hypergeometric series
\begin{equation}
\label{eq:solHyperEq}
f_i=K^{1-\beta_i}~ {}_nF_{n-1}(1+\alpha_1-\beta_i,\dots,1+\alpha_n-\beta_i;1+\beta_1-\beta_i,\check{\dots} ,1+\beta_n-\beta_i; K)\,,
\end{equation}
where $\check{}$ denotes omission of $1+\beta_i-\beta_i$ and the generalized hypergeometric series ${}_nF_{n-1}$ is defined by the formula
\begin{equation}
\label{eq:HyperSeries}
{}_nF_{n-1}(\alpha_1,\dots,\alpha_n;\beta_1,\dots, \beta_{n-1}; z)= \sum_{m\geq 0}^{\infty} \dfrac{\prod_{j=1}^{n}(\alpha_j)_m}{\prod_{j=1}^{n-1}(\beta_k)_m} \dfrac{z^m}{m!}\,,
\end{equation}
with $m\in \mathbb{N}$, $\alpha,\beta \in \mathbb{C}$ and $(a)_m$ is the Pochhammer symbol defined by
\begin{equation}
(a)_m = \begin{cases}
1 \,, ~~~~~~~~~~~~~~~~~~~~~~~~~~~~~~\qquad m=0 \,, \\
a(a+1)\dots(a+m-1) \,,\ \qquad m\geq 1\,.
\end{cases}
\end{equation}
For the cases of $d>3$, MLDE can only be solved recursively to obtain the $q$-expansion of the corresponding VVMFs.

\subsection{\label{sec:1-d VVMF}One-dimensional irreducible VVMFs}

From the relations $S^4=(ST)^3=1$ satisfied by the modular generators $S$ and $T$, we know that the $SL(2,\mathbb{Z})$ group has 12 irreps denoted as $\mathbf{1}_p$ with
\begin{equation}
\label{eq:SL2Zsinglets}
\mathbf{1}_p:~~~~~	\rho_{\mathbf{1}_p}(S)=i^p\,,\qquad \rho_{\mathbf{1}_p}(T)=e^{\frac{i\pi}{6} p}\,,
\end{equation}
where $p=0,1,\dots,11$. Then one can read off the exponent matrix $L=p/12$ for $\mathbf{1}_p$.

From the previous general result, the 1-d VVMFs space $\mathcal{M}(\rho)$ is the free $\mathcal{M}(\mathbf{1})$-module of rank 1. Let $F$ denote the basis of $\mathcal{M}(\rho)$ with the minimal weight $k_0$, thus $D_{k_0}F$ must be a linearly dependent on $F$, i.e. $D_{k_0} F= g F$ with coefficients $g\in \mathcal{M}(\mathbf{1})$.
Since there are no holomorphic modular forms of weight 2 in $\mathcal{M}(\mathbf{1})$,  it follows that $g=0$.
Therefore, the 1-d VVMFs satisfies the  following 1st order modular linear differential equation:
\begin{equation}
\label{eq:1stMLDE}
D_{k_0} F = 0 \,.
\end{equation}
From the Eq.~\eqref{eq:Ramanujan} and Eq.~\eqref{eq:j-Delta}, we can know that $D_{12} \Delta=0$. Because $\Delta=\eta^{24}$ where $\eta(\tau)$ is the Dedekind eta function\footnote{It is defined by $\eta(\tau)=q^{1/24}\prod_{n=1}^\infty \left(1-q^n \right)$ with $q=e^{2\pi i\tau}$.} and using the Leibniz law~\eqref{eq:Leibniz law}, we can obtain the important result $D_1 \eta^2=0$. This means that the eta products $\eta^{2k_0}$ is a solution of Eq.~\eqref{eq:1stMLDE}:
\begin{equation}
\label{eq:1stMLDEsolution}
F = \eta^{2k_0}\,.
\end{equation}
Furthermore, one can check that $\eta^{2k_0}(\tau)$ satisfies the following transformation form
\begin{align}
\label{eq:eta2p_tra}
\nonumber&\eta^{2k_0}(S\tau)=(-\tau)^{k_0} i^{k_0}\eta^{2k_0}(\tau) = (-\tau)^{k_0} \rho_{\mathbf{1}_{k_0}}(S)\eta^{2k_0}(\tau),\\
&\eta^{2k_0}(T\tau)=e^{\frac{i\pi}{6}k_0}\eta^{2k_0}(\tau)=\rho_{\mathbf{1}_{k_0}}(T)\eta^{2k_0}(\tau)\,.
\end{align}
Therefore $\eta^{2k_0}(\tau)$ is the weight $k_0$ one-dimensional VVMFs of $SL(2,\mathbb{Z})$ in the irrep $\rho_{\mathbf{1}_{k_0}}$. As a result, the one-dimensional VVMFs module is
\begin{equation}
\mathcal{M}(\mathbf{1}_p)=\mathbb{C}[E_4,E_6]\eta^{2p}\,,\qquad p=0,1,\dots,11\,.
\end{equation}

\subsection{\label{sec:2-d VVMF}Two-dimensional irreducible VVMFs}

All the 2-d irreducible unitary representations of $SL(2,\mathbb{Z})$ with finite image have been fully obtained in Ref.~\cite{mason20082}.
In the $T$-diagonal basis, the representation matrix of $T$ can be generally parameterized as
\begin{equation}
\label{eq:DiagT-2d}
\rho(T)=\begin{pmatrix}
 e^{2\pi i r_1} & 0 \\
 0 & e^{2\pi i r_2}
\end{pmatrix}\,,  \qquad 0\leq r_1,r_2 < 1\,.
\end{equation}
A striking feature is that these 2-d irreps are completely specified by the parameters $r_1$ and $r_2$. The representation matrix $\rho(S)$ is uniquely determined by $r_{1,2}$ up to a possible similar transformation, and it can be parameterized as~\cite{mason20082}
\begin{equation}
\label{eq:rhoS-2d}\rho(S)=\frac{1}{\lambda_1\lambda_2(\lambda_1-\lambda_2)}\begin{pmatrix}
1 & \sqrt{-(\lambda_1\lambda_2)^5(\lambda_1-\lambda_2)^2-1} \\
	\sqrt{-(\lambda_1\lambda_2)^5(\lambda_1-\lambda_2)^2-1} & -1
\end{pmatrix}\,,
\end{equation}
with $\lambda_{1,2}=e^{2\pi i r_{1,2}}$. The representation $\rho$ is irreducible and its determinant $\det\rho$ is an even 1-d representation of $SL(2,\mathbb{Z})$, thus the parameters $r_{1, 2}$ should satisfy~\cite{mason20082}
\begin{equation}
r_1+r_2 \in \frac{1}{6}\mathbb{Z}\,, \quad r_1-r_2 \notin \mathbb{Z}\,,~ \mathbb{Z}\pm\frac{1}{6}\,.
\end{equation}
We will denote the 2-d irreps as $\mathbf{2}_{(r_1,r_2)}$ whose exponent matrix $L$ reads as $L=\texttt{diag}(r_1,r_2)$. The modular group $SL(2,\mathbb{Z})$ has 54 inequivalent irreps with finite image, and the corresponding values of the unordered pair $(r_1,r_2)$ are summarized in the table~\ref{tab:2-d irrep}.

From the general theory of VVMFs in section~\ref{sec:VVMF}, we know that the 2-d VVMFs space $\mathcal{M}(\rho)$ is a free $\mathcal{M}(\mathbf{1})$-module of rank 2, and it has the free basis $\{F,D_{k_0}F\}$ with the minimal weight $k_0=6\text{Tr}(L)-1=6(r_1+r_2)-1$. Because $D^2_{k_0}F$ is a weight $k_0+4$ VVMF, it can be written as the linear combination of the free basis over $\mathcal{M}(\mathbf{1})$. Thus, we have
\begin{equation}
\label{eq:2thMLDE}
(D^2_{k_0} +a E_4) F=0 \,,
\end{equation}
with $a$ is a complex number. This is just a second order MLDE.

Since $D_{k_0}(\eta^{2k_0})=0$ as shown in section~\ref{sec:1-d VVMF}, we can always reduce the weight $k_0$ MLDE to weight zero MLDE by rescaling $F$ to $\widetilde{F}=\eta^{-2k_0}F$:
\begin{equation}
\label{eq:2th0MLDE}
(D^2_0 +a E_4) \widetilde{F}=0 \,,
\end{equation}
As shown in Ref.~\cite{franc2016hypergeometric}, by introducing the notations
\begin{align}
\nonumber& A=E_6/E_4\,,\qquad K(\tau)=1728/j(\tau)\,, \\
& \theta=q\frac{d}{dq} \,,~~~~\qquad \theta_K=K\frac{d}{dK}\,,
\end{align}
where $K(\tau)=1728q(1-744q+\dots)$ is a local parameter, the MLDE~\eqref{eq:2th0MLDE} can be written into a hypergeometric differential equation:
\begin{equation}
\label{eq:2thHyperDE}
\left(\theta_K^2-\frac{2K+1}{6(1-K)}\theta_K+\frac{a}{1-K} \right)\widetilde{F}=0\,.
\end{equation}
From the general $q$-expansion of Eq.~\eqref{eq:q-form}, we know  $\widetilde{F}$ can be expanded in a $K$-series,
\begin{equation}
\widetilde{F}=\eta^{-2k_0}F=\begin{pmatrix}
K^{r_1-\frac{k_0}{12}}\sum_{n\geq 0}a_1(n)K^n \\  K^{r_2-\frac{k_0}{12}}\sum_{n\geq 0}a_2(n)K^n
\end{pmatrix}\,.
\end{equation}
Considering the leading term of Eq.~\eqref{eq:2thHyperDE} in $K$, we can fix the values of $k_0$ and $a$,
\begin{equation}
k_0=6(r_1+r_2)-1\,,\qquad a=\left(r_1-\frac{k_0}{12}\right)\left(r_2-\frac{k_0}{12}\right)\,.
\end{equation}
Then the differential equation~\eqref{eq:2thHyperDE} can be written as the generalized hypergeometric equations of Eq.~\eqref{eq:generalHyperDE} with
\begin{equation}
\beta_1=\frac{r_2-r_1}{2}+\frac{11}{12}\,,\quad \beta_2=\frac{r_1-r_2}{2}+\frac{11}{12}\,,\quad
\alpha_1=0\,,\quad \alpha_2=\frac{1}{3}\,.
\end{equation}
Therefore the components of $\widetilde{F}$ as the solutions of Eq.~\eqref{eq:2th0MLDE} are hypergeometric series of type ${}_2F_1$, as shown in Eq.~\eqref{eq:solHyperEq}, and the 2-d VVMF $F$ is of the following form
\begin{equation}
\label{eq:2dVVMF}
F(\tau)=
\begin{pmatrix}
~~\eta^{12(r_1+r_2)-2}K^{\frac{6(r_1-r_2)+1}{12}}~ {}_2F_1(\frac{6(r_1-r_2)+1}{12},\frac{6(r_1-r_2)+5}{12};r_1-r_2+1;K) \\
C \eta^{12(r_1+r_2)-2}K^{\frac{6(r_2-r_1)+1}{12}}~ {}_2F_1(\frac{6(r_2-r_1)+1}{12},\frac{6(r_2-r_1)+5}{12};r_2-r_1+1;K) \\
\end{pmatrix}
\end{equation}
up to an overall irrelevant constant. The coefficient $C$ depends on the explicit form of the representation matrix $\rho(S)$, and its value can be fixed to satisfy the condition Eq.~\eqref{eq:defVVMF} of VVMF. Hence the module of two-dimensional VVMFs is given by
\begin{equation}
\mathcal{M}(\mathbf{2}_{(r_1,r_2)})=\mathbb{C}[E_4,E_6]F\oplus \mathbb{C}[E_4,E_6]D_{k_0}F\,.
\end{equation}

\subsection{\label{sec:3-d VVMF}Three-dimensional irreducible VVMFs}
The 3-d irreducible unitary representations of $SL(2,\mathbb{Z})$ with finite image have not been fully studied as far as we know, a complete classification is not available in the literature at present. In the $T$ diagonal basis, the 3-d representation matrix $\rho(T)$ can be parameterized as
\begin{equation}
	\label{eq:DiagT}
	\rho(T)=\begin{pmatrix}
		e^{2\pi i r_1} & 0 & 0 \\
		0 & e^{2\pi i r_2} &  0 \\
		0 & 0 & e^{2\pi i r_3}
	\end{pmatrix}\,,  \qquad 0\leq r_1,r_2,r_3 < 1\,,
\end{equation}
which implies the exponent matrix $L=\texttt{diag}(r_1,r_2,r_3)$. See Refs.~\cite{tuba2001representations,Altarelli:2005yx} for alternative parameterizations for both $\rho(S)$ and $\rho(T)$. Similar to the 2-d case, these 3-d irreps are uniquely determined by the parameters $r_1,r_2$ and $r_3$~\cite{tuba2001representations}. Consequently we denote these 3-d irreps by $\mathbf{3}_{(r_1,r_2,r_3)}$. The determinant of $\rho$ forms a 1-d irrep of $SL(2,\mathbb{Z})$ in Eq.~\eqref{eq:SL2Zsinglets}, therefore the parameters $r_{1,2,3}$ have to fulfill the following constraint:
\begin{equation}
r_1+r_2+r_3 \in \frac{1}{12}\mathbb{Z} \,.
\end{equation}
The space $\mathcal{M}(\rho)$ of 3-d VVMFs is a rank 3 module over $\mathcal{M}(\mathbf{1})$ with the free basis $\{F,D_{k_0}F,D^2_{k_0}F\}$ and the minimal weight $k_0=4\text{Tr}(L)-2=4(r_1+r_2+r_3)-2$. As shown in Eq.~\eqref{eq:example3dMLDE}, $D^3_{k_0}F$ can be expressed by the free basis and consequently the minimal weight VVMF $F$ satisfies the following 3rd order MLDE
\begin{equation}
\label{eq:3thMLDE}
(D^3_{k_0} +a E_4 D_{k_0}+bE_6) F=0 \,,
\end{equation}
where $a$ and $b$ are two complex numbers. Reparameterizing this equation via the local parameter $K$, one obtains
\begin{equation}
\label{eq:3thHyperDE}
\left(\theta_K^3-\frac{2K+1}{2(1-K)}\theta^2_K+\frac{18a+1-4K}{18(1-K)}\theta_K +\frac{b}{1-K}\right) \widetilde{F}=0\,.
\end{equation}
where $\widetilde{F}=\eta^{-2k_0}F$. In the same fashion as section~\ref{sec:2-d VVMF}, we find the values of $k_0$, $a$ and $b$ are
\begin{align}
\nonumber&
k_0=4(r_1+r_2+r_3)-2\,,\\ \nonumber&a=\left(r_1-\frac{k_0}{12}\right)\left(r_2-\frac{k_0}{12}\right)+\left(r_1-\frac{k_0}{12}\right)\left(r_3-\frac{k_0}{12}\right)+\left(r_2-\frac{k_0}{12}\right)\left(r_3-\frac{k_0}{12}\right)-\frac{1}{18}\,,\\
&b=-\left(r_1-\frac{k_0}{12}\right)\left(r_2-\frac{k_0}{12}\right)\left(r_3-\frac{k_0}{12}\right)\,.
\end{align}
The differential equation~\eqref{eq:3thHyperDE} can immediately be expressed in the form of Eq.~\eqref{eq:generalHyperDE} with
\begin{align}
\nonumber&
\beta_1=\frac{5}{6}-\frac{1}{6}(4r_1-2r_2-2r_3)\,,\\
\nonumber&
\beta_2=\frac{5}{6}-\frac{1}{6}(4r_2-2r_3-2r_1)\,,\\
\nonumber&
\beta_3=\frac{5}{6}-\frac{1}{6}(4r_3-2r_1-2r_2)\,,\\
&\alpha_1=0\,,\quad \alpha_2=\frac{1}{3}\,,\quad \alpha_2=\frac{2}{3}\,.
\end{align}
Hence the VVMF $F$ is determined to be
\begin{equation}
\label{eq:3dVVMF}
F(\tau)=
\begin{pmatrix}
~~\eta^{8(r_1+r_2+r_3)-4}K^{\frac{a_1+1}{6}}~ {}_3F_2(\frac{a_1+1}{6},\frac{a_1+3}{6},\frac{a_1+5}{6};r_1-r_2+1,r_1-r_3+1;K) \\
C_1\eta^{8(r_1+r_2+r_3)-4}K^{\frac{a_2+1}{6}}~ {}_3F_2(\frac{a_2+1}{6},\frac{a_2+3}{6},\frac{a_2+5}{6};r_2-r_3+1,r_2-r_1+1;K) \\
C_2\eta^{8(r_1+r_2+r_3)-4}K^{\frac{a_3+1}{6}}~ {}_3F_2(\frac{a_3+1}{6},\frac{a_3+3}{6},\frac{a_3+5}{6};r_3-r_1+1,r_3-r_2+1;K)
\end{pmatrix}\,,
\end{equation}
with $a_1=4r_1-2r_2-2r_3,a_2=4r_2-2r_1-2r_3$ and $a_3=4r_3-2r_1-2r_2$. The constants $C_{1,2}$ are fixed by the representation matrix of $\rho(S)$ in the $T$-diagonal basis. The 3-d VVMFs module is of the form
\begin{equation}
\mathcal{M}(\mathbf{3}_{(r_1,r_2,r_3)})=\mathbb{C}[E_4,E_6]F\oplus \mathbb{C}[E_4,E_6]D_{k_0}F\oplus \mathbb{C}[E_4,E_6]D^2_{k_0}F\,.
\end{equation}
For high dimensional irreducible VVMFs, the MLDE cannot be converted to generalize hypergeometric equation, and can only be solved recursively. Some results of 4-d irreducible VVMFs can be found in Ref.~\cite{franc2020constructions}, we shall not discuss these cases further in the present work.

\section{\label{sec:framework}The general finite modular groups and modular invariant theory }

The fundamental theorem of homomorphism implies that $\ker(\rho)$ is a normal subgroup of $SL(2,\mathbb{Z})$, and the image $\text{Im}(\rho)\cong \Gamma/\ker(\rho)$ forms a discrete finite group since the irrep $\rho$ has finite image. For the modular transformation $\gamma\in\ker(\rho)$ in Eq.~\eqref{eq:defVVMF}, we have $\rho(\gamma)=\mathbb{1}_{d}$ and consequently each component of VVMF $Y(\tau)$ in the representation $\rho$ is scalar modular form of $\ker(\rho)$. In the paradigm of modular flavor symmetry~\cite{Feruglio:2017spp}, $\ker(\rho)$ was only restricted to principle congruence subgroups $\Gamma(N)$. The image $\text{Im}(\rho)$ is the homogeneous finite modular group $\Gamma'_N=\Gamma/\Gamma(N)$ and it is taken as the flavor symmetry to address the flavor structure of quarks and leptons. From the view of VVMF, the kernel $\ker(\rho)$ can be a general normal subgroup of $SL(2,\mathbb{Z})$ instead of $\Gamma(N)$, and accordingly the finite modular group does not necessarily have to be $\Gamma'_N$. Thus the framework of modular invariance can be extended significantly by considering VVMFs, and we have more possible choices for the finite modular groups to construct modular invariant models. Obviously the irreps of the finite modular groups can be lifted to the irreps of $SL(2,\mathbb{Z})$. The finite image irreps of $SL(2,\mathbb{Z})$ can be obtained from these (infinite many) general finite modular groups.

The classification of normal subgroups of $PSL(2,\mathbb{Z})$ was first studied in the 1960s~\cite{newman1964,newman1967}, and the normal subgroups of index up to $1500$ are given in Ref.~\cite{conder2006normal}. The normal subgroups of $SL(2,\mathbb{Z})$ have not been fully discussed as those of $PSL(2,\mathbb{Z})$. Some methods for finding normal subgroups of small index in finitely-presented groups have been developed in computational group theory~\cite{holt2005handbook}, and an algorithm has been implemented as the command $\texttt{LowIndexNormalSubgroups}$ in the package MAGMA~\cite{Magma:2021}. The normal subgroups of $SL(2,\mathbb{Z})$ with small index can be easily obtained with the help of MAGMA. In table~\ref{tab:NorSubgroupSL2Z} we list the normal subgroups of $SL(2,\mathbb{Z})$ with index $\leq 72$ as well as the corresponding finite modular group, where we omit the few cases for which the finite modular group is abelian. Besides the known principal congruence subgroups $\Gamma(N)$
as well as the inhomogeneous and homogeneous finite modular groups, we see other normal subgroups of $SL(2,\mathbb{Z})$. Additional relators are elements of $\ker(\rho)$, when they are added to the $SL(2,\mathbb{Z})$ presentation relations $S^4=(ST)^3=1$,  $S^2 T=T S^2$ in Eq.~\eqref{eq:SL2Zrelation}, the finite modular groups $\text{Im}(\rho)\cong \Gamma/\ker(\rho)$ are produced. Moreover, if the element $S^2\in \ker(\rho)$, we have $\rho(S^2)=\mathbb{1}_d$. Then the corresponding finite modular group has only even representations, and the VVMFs in these irreps must be of even weights.

\begin{table}[t!]
\centering
\begin{tabular}{|c|c|c||c|c|c|c|c|c|c|c|c|c|c|c|c|}\hline\hline
\multicolumn{3}{|c||}{Normal subgroups $\ker(\rho)$} & \multicolumn{2}{c|}{Finite modular groups $\Gamma/\ker(\rho)\cong\text{Im}(\rho)$} \\ \hline
Index& Label & Additional relators & Group structure  & \texttt{GAP} Id \\
\hline		
6 & $\Gamma(2)$ & $T^2$ & $S_3$ & $[6,1]$  \\ \hline	
\multirow{2}{*}{12} & $-$ & $S^2 T^2$ &  $ Z_3\rtimes Z_4\cong 2D_{3}$ & $[12,1]$  \\ \cline{2-5}
			
 & $\pm\Gamma(3)$ & $S^2, T^3$ & $A_4$ & $[12,3]$  \\ \hline
			
18 & $-$  & $ST^{-2}ST^2$ & $S_3\times Z_3$ & $[18,3]$  \\ \hline
			
\multirow{4}{*}{24} & $\Gamma(3)$ & $T^3$ & \multirow{2}{*}{$T'$} & \multirow{2}{*}{$[24,3]$}  \\ \cline{2-3}
			
	& $-$ & $S^2T^3$ &  &   \\  \cline{2-5}
			
	& $\pm\Gamma(4)$ & $S^2,T^4$ & $S_4$ & $[24,12]$  \\ \cline{2-5}
			
    & $-$ &  $S^2,(ST^{-1}ST)^2$ & $A_4\times Z_2$ & $[24,13]$  \\ \hline

 36	& $-$ &  $S^3T^{-2}ST^2$ & $(Z_3\rtimes Z_4) \times Z_3$ & $[36,6]$  \\\hline		

\multirow{2}{*}{42} & $-$ & $T^6,(ST^{-1}S)^2TST^{-1}ST^2$ & \multirow{2}{*}{$Z_7 \rtimes Z_6$} & \multirow{2}{*}{$[42,1]$}  \\ \cline{2-3}
			
			& $-$ & $T^6,ST^{-1}ST(ST^{-1}S)^2T^2$ &  &   \\ \hline
			
\multirow{6}{*}{48} & $-$ & $S^2T^4$ & $2O$ & $[48,28]$  \\ \cline{2-5}

& $-$ & $T^8,ST^4ST^{-4}$ & $GL(2,3)$ & $[48,29]$  \\ \cline{2-5}

& $\Gamma(4)$ & $T^4$ & $A_4\rtimes Z_4\cong S'_4$ & $[48,30]$  \\ \cline{2-5}

& $-$ & $(ST^{-1}ST)^2$ & $A_4\times Z_4$ & $[48,31]$  \\ \cline{2-5} 		
	
& $-$ & $S^2(ST^{-1}ST)^2$ & $T'\times Z_2$ & $[48,32]$  \\ \cline{2-5}
			
& $-$ & $T^{12},ST^3ST^{-3}$ & $((Z_4\times Z_2)\rtimes Z_2) \rtimes Z_3$ & $[48,33]$  \\ \hline

 54 & $-$ & $T^{6},(ST^{-1}ST)^3$ & $(Z_3\times Z_3)\rtimes Z_6 $ & $[54,5]$  \\ \hline

 60 & $\pm\Gamma(5)$ & $S^2, T^{5}$ & $A_5$ & $[60,5]$  \\ \hline

\multirow{2}{*}{72} & $-$ & $T^{12},ST^4ST^{-4}$ & $S_4\times Z_3$ & $[72,42]$  \\ \cline{2-5}
& $\pm\Gamma(6)$ & $S^2,T^{6},(ST^{-1}STST^{-1}S)^2T^2$ & $A_4\times S_3$ & $[72,44]$  \\ \hline\hline
\end{tabular}
\caption{\label{tab:NorSubgroupSL2Z} The normal subgroups of $SL(2,\mathbb{Z})$ with index less than $78$ and the corresponding finite modular groups. It is defined $\pm\Gamma(N)=\left\{\pm\gamma, \gamma\in\Gamma(N)\right\}$. Note that $2D_{3}$ is the binary dihedral group of order 12 and $2O$ is the binary octahedral group which is the Schur cover of permutation group $S_4$ of type ``$-$''. Additional relators are some elements of $\ker(\rho)$, the quotient group $\Gamma/\ker(\rho)$ is produced when the relations ``Additional relators =1'' together with $S^4=(ST)^3=1$ and $S^2 T=T S^2$ in Eq.~\eqref{eq:SL2Zrelation} are imposed.}
\end{table}

In the following, we generalize the modular invariant theory~\cite{Feruglio:2017spp} by considering the general finite modular groups denoted as $\mathcal{G}_f$ rather than $\Gamma'_N$ and $\Gamma_N$. The theory depends on a set of chiral supermultiplets $\Phi_I$ and the modulus superfield $\tau$ which transform under the modular group $\Gamma$ as
\begin{equation}
\label{eq:modularTrs_Phi}
\begin{cases}
\tau\to \gamma\tau=\dfrac{a\tau+b}{c\tau+d}\,,\\[0.1in]
\Phi_I\to (c\tau+d)^{-k_I}\rho_I(\gamma)\Phi_I\,,
\end{cases}
\qquad \gamma=\begin{pmatrix}
a & b \\ c & d
\end{pmatrix} \in \Gamma\,,
\end{equation}
where $-k_I$ is the modular weight of matter field $\Phi_I$, and $\rho_I(\gamma)$ is the unitary irreducible representation of $\mathcal{G}_f$ which is the quotient between $\Gamma$ and its certain normal subgroups. $\mathcal{G}_f$ acts as a flavor symmetry group and it is kept fixed in the construction. We choose a minimal form of the Kahler potential~\cite{Feruglio:2017spp}
\begin{equation}
\label{eq:minKahler}
\mathcal{K}(\Phi_I,\bar{\Phi}_I; \tau,\bar{\tau}) =-h\Lambda^2 \log(-i\tau+i\bar\tau)+ \sum_I (-i\tau+i\bar\tau)^{-k_I} |\Phi_I|^2\,,
\end{equation}
where $h$ is a positive constant and $\Lambda$ is the cut-off scale of the theory. After the modulus $\tau$ acquires a nonzero vacuum expectation value, this Kahler potential gives rise to kinetic terms for both the matter fields and the modulus field $\tau$. Similar to the original modular flavor symmetry, the most general K\"ahler potential consistent with modular symmetry contains additional terms~\cite{Chen:2019ewa}. The superpotential $\mathcal{W}(\Phi_I,\tau)$ can be expanded in power series of the matter supermultiplets $\Phi_I$,
\begin{equation}
	\mathcal{W}(\Phi_I,\tau) =\sum_n Y_{I_1...I_n}(\tau)~ \Phi_{I_1}... \Phi_{I_n}\,.
\end{equation}
Clearly modular invariance of the superpotential requires the function $Y_{I_1...I_n}(\tau)$ should transform in the following way:
\begin{equation}
Y_{I_1...I_n}(\tau)\stackrel{\gamma}{\longrightarrow}Y_{I_1...I_n}(\gamma\tau)=(c\tau+d)^{k_Y}\rho_{Y}(\gamma)Y_{I_1...I_n}(\tau),~~~\gamma=\begin{pmatrix}
a &~ b \\ c &~ d
\end{pmatrix} \in \Gamma\,,
\end{equation}
where $\rho_Y$ is an irrep of $\mathcal{G}_f$, and $k_Y$ and $\rho_Y$ satisfy the conditions
\begin{eqnarray}
\nonumber&&k_Y=k_{I_1}+...+k_{I_n}\,,\\
&&\rho_Y\otimes\rho_{I_1}\otimes\cdots\otimes\rho_{I_n}\ni\bm{1}\,.
\end{eqnarray}
Hence $Y_{I_1...I_n}(\tau)$ is a VVMF of weight $k_Y$ and in the representation $\rho_Y$ of the finite modular group $\mathcal{G}_f$, and its explicit forms can be obtained by solving the MLDE, as shown in section~\ref{sec:resultVVMF}. The number of VVMFs in a given irrep $\rho_Y$ generally increases with the modular weight $k_Y$. Consequently the larger is $k_{I_1}+...+k_{I_n}$, more modular invariant terms are involved in $\mathcal{W}$. However, the module $\mathcal{M}(\rho_Y)$ is finitely generated by $d=\dim\rho_Y$ independent bases denoted by $\{Y_1,\dots, Y_d\}$, thus $Y_{I_1...I_n}(\tau)$ can be written uniformly in the form of $Y^{(k_Y)}_{\rho_Y}=\alpha_1 Y_1+\dots +\alpha_d Y_d$ with $\alpha_i\in\mathbb{C}[E_4,E_6]$, note that $\alpha_i$ are polynomials of $E_4$ and $E_6$. Some $\alpha_i$ for vanishing for small value of $k_Y$, nevertheless there are at most $d$ independent terms. This implies that the superpotential is strongly constrained by the modular symmetry, and it can only take a finite number of possible forms for a given $\mathcal{G}_f$, although there are infinite possible weight and representation assignments for the matter fields. See Appendix~\ref{app:finiteness} for explanation.

\section{\label{sec:example}New modular invariant lepton models }

In this section, we shall apply the formalism of section~\ref{sec:framework} to construct models of lepton masses and mixings. We shall be concerned with the finite modular groups beyond $\Gamma_N$ and $\Gamma'_N$. For illustration, we will consider $A_4\times Z_2$ and $GL(2,3)$ as the finite flavor groups and two benchmark models will be constructed without introducing any flavon. Notice that these two finite groups as flavor symmetry have not been discussed in the literature. As shown in table~\ref{tab:NorSubgroupSL2Z}, it is remarkable that there are some other interesting finite modular groups which could be helpful to understand the flavor puzzle, such as the binary dihedral group $2D_3$ and the binary octahedral group $2O$ which are left for future work\footnote{The group $2D_3$ has four 1-d irreps and two 2-d irreps. The group $2O$ is of order 48 and it is the Schur cover of the permutation group $S_4$ of type ``$-$''. It has two 1-d irreps, three 2-d irreps, two 3-d irreps and one 4-d irrep, the three generations of lepton doublets could be embedded into a triplet of $2O$ as usual.}.

\subsection{\label{subsec:model1}Model 1 based on finite modular group $A_4\times Z_2$}

The group $A_4\times Z_2$ is the direct product of the tetrahedral group $A_4$ and the cyclic group $Z_2$, and it can be generated by two elements $S$ and $T$ obeying the relations
\begin{equation}
\label{eq:multiplication-rule-A4xZ2}S^2=(ST)^3=(ST^{-1}ST)^2=1 \,.
\end{equation}
The 24 group elements can be expressed in terms of $S$ and $T$ as : $1$, $S$, $T^{-1}$, $T$, $ST^{-1}$, $ST$, $T^{-1}S$, $T^{-2}$, $TS$, $T^2$, $ST^{-1}S$, $ST^{-2}$, $STS$, $ST^2$, $T^{-1}ST$, $T^{-2}S$, $TST^{-1}$, $T^2S$, $T^3$, $ST^{-2}S$, $ST^2S$, $T^{-1}STS$, $TST^{-1}S$, $T^3S$. This group has eight inequivalent irreps:
six 1-d irreducible representations denoted by $\bm{1}_0,\bm{1}_0',\bm{1}_0'',\bm{1}_1,\bm{1}_1',\bm{1}_1''$
and two 3-d irreducible representations denoted by
$\bm{3}_0,\bm{3}_1$. It is immediate to see that 1-d unitary representations are given by
\begin{align}
\nonumber& \bm{1}_i ~:~~ S=(-1)^i\,,~~~~~\quad  T= (-1)^i \,,\\
\nonumber& \bm{1}'_i ~:~~ S=(-1)^i\,,~~~~~\quad  T = (-1)^i \omega\,,\\
& \bm{1}''_i ~:~~ S=(-1)^i\,,~~~~~\quad T = (-1)^i \omega^2 \,,
\end{align}
where $i=0,1$ and $\omega=e^{2\pi i /3}$. We see that these six 1-d representations coincide with the $SL(2,\mathbb{Z})$ singlet representations $\mathbf{1}_0$, $\mathbf{1}_2$, $\mathbf{1}_4$, $\mathbf{1}_6$, $\mathbf{1}_8$, $\mathbf{1}_{10}$ and $\mathbf{1}_{12}$ in Eq.~\eqref{eq:SL2Zsinglets}. The 3-d irreps $\bm{3}_0$ and $\bm{3}_1$ in the $T$ diagonal basis is given by,
\begin{equation}
\bm{3}_i ~:~~S=\frac{(-1)^i}{3} \begin{pmatrix}
-1 & 2 & 2 \\ 2 & -1 & 2 \\ 2 & 2 & -1
\end{pmatrix}
\,,\,\qquad T = (-1)^i \begin{pmatrix}
1 & 0 & 0 \\ 0 & \omega & 0 \\ 0 & 0 & \omega^2
\end{pmatrix}\,,
\end{equation}
which are exactly the triplet representation $\mathbf{3}_{(0,\frac{1}{3},\frac{2}{3})}$ and $\mathbf{3}_{(\frac{1}{2},\frac{5}{6},\frac{1}{6})}$ of $SL(2,\mathbb{Z})$.
The relevant multiplication rules are
\begin{eqnarray}
\nonumber&&\bm{1}'_i \otimes \bm{1}'_j = \bm{1}''_{\langle i+j\rangle},~~~\bm{1}''_i \otimes \bm{1}''_j = \bm{1}'_{\langle i+j\rangle},~~~\bm{1}'_i \otimes \bm{1}''_j = \bm{1}_{\langle i+j\rangle}\,,\\
&&\bm{3}_i \otimes \bm{3}_j =\bm{1}_{\langle i+j\rangle} \oplus \bm{1'}_{\langle i+j\rangle} \oplus \bm{1''}_{\langle i+j\rangle}
\oplus \bm{3}_{\langle i+j\rangle,S} \oplus \bm{3}_{\langle i+j\rangle,A}
\end{eqnarray}
where $\langle i+j\rangle\equiv i+j~\texttt{mod}\;2$, and $\bm{3}_{\langle i+j\rangle, S}$ and $\bm{3}_{\langle i+j\rangle, A}$ stand for the symmetric and the antisymmetric triplet combinations respectively. The decomposition of the tensor product of two triplets $\bm{\alpha}=(\alpha_1, \alpha_2, \alpha_3)^{T}\sim\mathbf{3}_i$ and $\bm{\beta}=(\beta_1, \beta_2, \beta_3)^{T}\sim\mathbf{3}_j$ is
\begin{eqnarray}
\nonumber&&\left(\bm{\alpha} \otimes \bm{\beta} \right)_{\bm{1}_{\langle i+j\rangle}}=\alpha_1\beta_1+\alpha_2\beta_3+\alpha_3\beta_2 \,,\\
\nonumber&&\left(\bm{\alpha} \otimes \bm{\beta} \right)_{\bm{1'}_{\langle i+j\rangle} }=\alpha_3\beta_3+\alpha_1\beta_2+\alpha_2\beta_1 \,, \\
\nonumber&&\left(\bm{\alpha} \otimes \bm{\beta} \right)_{\bm{1''}_{\langle i+j\rangle}}=\alpha_2\beta_2+\alpha_1\beta_3+\alpha_3\beta_1 \,, \\
\nonumber&&\left(\bm{\alpha} \otimes \bm{\beta} \right)_{\bm{3}_{\langle i+j\rangle,S}}=\begin{pmatrix}2\alpha_1\beta_1-\alpha_2\beta_3-\alpha_3\beta_2 \\
2\alpha_3\beta_3-\alpha_1\beta_2-\alpha_2\beta_1 \\
2\alpha_2\beta_2-\alpha_1\beta_3-\alpha_3\beta_1 \\
\end{pmatrix}\,,\\
&&\left(\bm{\alpha} \otimes \bm{\beta} \right)_{\bm{3}_{\langle i+j\rangle,A}}=\begin{pmatrix}\alpha_3\beta_2-\alpha_2\beta_3 \\
\alpha_2\beta_1-\alpha_1\beta_2 \\
\alpha_1\beta_3-\alpha_3\beta_1 \\
\end{pmatrix}\,.\label{eq:tensor-product-A4xZ2}
\end{eqnarray}

There are no odd weight modular forms in the representation of $A_4\times Z_2$, because $S^2=1$ shown in Eq.~\eqref{eq:multiplication-rule-A4xZ2} and consequently $\rho(S^2)=\mathbb{1}_d$. From the table~\ref{tab:IrrepProfile} or the dimension formula of VVMF in Eq.~\eqref{eq:dim-formula}, we can deduce that there are two modular multiplets in the representation $\bm{1}''_1$ and $\bm{3}_0$ at weight 2. Their concrete forms can be directly obtained from the general results of Eqs.~(\ref{eq:1stMLDEsolution}, \ref{eq:3dVVMF}) by considering the $S$ transformations, i.e.,
\begin{equation}
\label{eq:wt2A4Z2VVMF} Y^{(2)}_{\bm{1}_1''}(\tau)= \eta^4(\tau),\quad ~~~Y^{(2)}_{\bm{3}_0}(\tau)=\begin{pmatrix}
\eta^4(\tau)(\frac{K(\tau)}{1728})^{-\frac{1}{6}}~ {}_3F_2(-\frac{1}{6},\frac{1}{6},\frac{1}{2};\frac{2}{3},\frac{1}{3};K(\tau)) \\
-6 \eta^4(\tau)(\frac{K(\tau)}{1728})^{\frac{1}{6}}~ {}_3F_2(\frac{1}{6},\frac{1}{2},\frac{5}{6};\frac{4}{3},\frac{2}{3};K(\tau)) \\
-18 \eta^4(\tau)(\frac{K(\tau)}{1728})^{\frac{1}{2}}~ {}_3F_2(\frac{1}{2},\frac{5}{6},\frac{7}{6};\frac{5}{3},\frac{4}{3};K(\tau))
\end{pmatrix}\,,
\end{equation}
Their $q$-expansions can be read out as follows,
\begin{align}
\label{eq:q-wt2A4Z2VVMF}
\nonumber&Y^{(2)}_{\bm{1}_1''}(\tau)= q^{1/6} (1 - 4 q + 2 q^2 + 8 q^3 - 5 q^4 - 4 q^5 - 10 q^6+\dots)\,,\\
&Y^{(2)}_{\bm{3}_0}(\tau)=\begin{pmatrix}
 1 + 12 q + 36 q^2 + 12 q^3 + 84 q^4 + 72 q^5 + 36 q^6 + \dots \\
-6 q^{1/3} (1 + 7 q + 8 q^2 + 18 q^3 + 14 q^4 + 31 q^5 + 20 q^6 + \dots)\\
-18 q^{2/3} (1 + 2 q + 5 q^2 + 4 q^3 + 8 q^4 + 6 q^5 + 14 q^6+\dots )
\end{pmatrix}\,.
\end{align}
We see $Y^{(2)}_{\bm{3}_0}(\tau)$ is exactly the triplet modular form of $\Gamma(3)$~\cite{Feruglio:2017spp}, and the over factor $q^{1/6}$ in $Y^{(2)}_{\bm{1}_1''}$ provides new ingredient to explain the mass hierarchy of charged leptons and quarks. The higher weight modular multiplets can be built from the tensor product of $Y^{(2)}_{\bm{3}_0}(\tau)$ and $Y^{(2)}_{\bm{1}''_1}(\tau)$ in eq.~\eqref{eq:wt2A4Z2VVMF}. For example, at weight 4, we have four modular multiplets
\begin{align}
\nonumber &Y^{(4)}_{\bm{1}_0}= (Y^{(2)}_{\bm{3}_0} Y^{(2)}_{\bm{3}_0})_{\bm{1}_0} = \big(Y^{(2)}_{\bm{3}_0,1}\big)^2+2Y^{(2)}_{\bm{3}_0,2}Y^{(2)}_{\bm{3}_0,3} \,, \\
\nonumber &Y^{(4)}_{\bm{1}'_0}= (Y^{(2)}_{\bm{3}_0} Y^{(2)}_{\bm{3}_0})_{\bm{1}'_0} = \big(Y^{(2)}_{\bm{3}_0,3}\big)^2+2Y^{(2)}_{\bm{3}_0,1}Y^{(2)}_{\bm{3}_0,2} \,, \\
\nonumber &Y^{(4)}_{\bm{3}_0}= \frac{1}{2}(Y^{(2)}_{\bm{3}_0} Y^{(2)}_{\bm{3}_0})_{\bm{3}_0}= \begin{pmatrix}
\big(Y^{(2)}_{\bm{3}_0,1}\big)^2-Y^{(2)}_{\bm{3}_0,2}Y^{(2)}_{\bm{3}_0,3} \\ \big(Y^{(2)}_{\bm{3}_0,3}\big)^2-Y^{(2)}_{\bm{3}_0,1}Y^{(2)}_{\bm{3}_0,2} \\ \big(Y^{(2)}_{\bm{3}_0,2}\big)^2-Y^{(2)}_{\bm{3}_0,1}Y^{(2)}_{\bm{3}_0,3}
\end{pmatrix}\,,\\
 &Y^{(4)}_{\bm{3}_1}= Y^{(2)}_{\bm{3}_0} Y^{(2)}_{\bm{1}''_1} = \begin{pmatrix}
Y^{(2)}_{\bm{3}_0,2}Y^{(2)}_{\bm{1}''_1} \\
Y^{(2)}_{\bm{3}_0,3}Y^{(2)}_{\bm{1}''_1} \\ Y^{(2)}_{\bm{3}_0,1}Y^{(2)}_{\bm{1}''_1}
\end{pmatrix}\,,\label{eq:wt4MF-A4xZ2}
\end{align}
where $Y^{(k)}_{\mathbf{r}, i}$ refers to the $i$th component of the VVMF $Y^{(k)}_{\mathbf{r}}$. The concrete forms of the higher weight modular multiplets in the representations of $A_4\times Z_2$ can be obtained in a similar way.

\subsubsection{Structure of the model}

The neutrino masses are described by type-I seesaw mechanism in this model. We introduce three right-handed neutrinos $N^c=(N^c_1, N^c_2, N^c_3)^T$ which furnish a  triplet $\bm{3}_1$ of $A_4\times Z_2$. The three generations of the left-handed leptons $L=(L_1,L_2,L_3)^T$ transform as a triplet $\bm{3}_0$ under $A_4\times Z_2$, and the right-handed charged leptons $E^c_{1,2,3}$ are assigned to singlet representations $\bm{1}_0,\bm{1}''_0,\bm{1}'_0$ respectively. The representation and weight assignments of the matter fields are summarized as follows:
\begin{align} \nonumber&\rho_{E^c}=\mathbf{1}''_0\oplus\mathbf{1}'_0\oplus\mathbf{1}''_1,~ \rho_{L} = \mathbf{3}_0,~\quad \rho_{N^c} = \mathbf{3}_1,~\rho_{H_u}=\rho_{H_d}=\mathbf{1}_0\,,\\
	&k_{E_{1,2,3}^c}=-1,1,1\,,~\quad k_{L}=3\,,~\quad k_{N^c}=1,~ \quad k_{H_u}=k_{H_d}=0\,.
\end{align}
The modular invariant superpotential for the charged lepton and neutrino masses is given by
\begin{align}
\nonumber &\hskip-0.1in \mathcal{W}_e =  \alpha \left( E_1^c L Y^{(2)}_{\bm{3}_0}\right)_{\bm{1}_0} H_d + \beta \left(E_2^c L Y^{(4)}_{\bm{3}_0}\right)_{\bm{1}_0} H_d + \gamma \left(E_3^c L Y^{(4)}_{\bm{3}_1}\right)_{\bm{1}_0} H_d\,, \\
 &\hskip-0.1in  \mathcal{W}_\nu =  g_1 \left( (N^c L)_{\bm{3}_1,S} Y^{(4)}_{\bm{3}_1}\right)_{\bm{1}_0} H_u +g_2 \left( (N^c L)_{\bm{3}_1,A} Y^{(4)}_{\bm{3}_1}\right)_{\bm{1}_0} H_u+ \Lambda \left( (N^c N^c)_{\bm{3}_0,S} Y^{(2)}_{\bm{3}_0}\right)_{\bm{1}_0}\,.
\end{align}
Using the decomposition rules in Eq.~\eqref{eq:tensor-product-A4xZ2}, we can read out the charged lepton and neutrino mass matrices as follows
\begin{align}
	\label{eq:A4Z2lept1}
	\nonumber&M_e=  \begin{pmatrix}
		\alpha Y^{(2)}_{\bm{3}_0,2} & \alpha Y^{(2)}_{\bm{3}_0,1} & \alpha Y^{(2)}_{\bm{3}_0,3} \\
		\beta Y^{(4)}_{\bm{3}_0,3} & \beta Y^{(4)}_{\bm{3}_0,2} & \beta Y^{(4)}_{\bm{3}_0,1} \\
		\gamma Y^{(4)}_{\bm{3}_1,2} & \gamma Y^{(4)}_{\bm{3}_1,1} & \gamma Y^{(4)}_{\bm{3}_1,3}
	\end{pmatrix}v_d  \,, \\
	\nonumber&M_D=\begin{pmatrix}
		2 g_1 Y^{(4)}_{\bm{3}_1,1}  & (-g_1+g_2) Y^{(4)}_{\bm{3}_1,3} & -(g_1+g_2)Y^{(4)}_{\bm{3}_1,2} \\
	    -(g_1+g_2)Y^{(4)}_{\bm{3}_1,3}  & 2 g_1 Y^{(4)}_{\bm{3}_1,2} & (-g_1+g_2) Y^{(4)}_{\bm{3}_1,1} \\
	  (-g_1+g_2) Y^{(4)}_{\bm{3}_1,2} & -(g_1+g_2)Y^{(4)}_{\bm{3}_1,1} &  2 g_1 Y^{(4)}_{\bm{3}_1,3}
	\end{pmatrix}v_u\,,\\
	&M_N= \Lambda \begin{pmatrix}
		2 Y^{(2)}_{\bm{3}_0,1} & -Y^{(2)}_{\bm{3}_0,3} & -Y^{(2)}_{\bm{3}_0,2}  \\
		-Y^{(2)}_{\bm{3}_0,3} & 2 Y^{(2)}_{\bm{3}_0,2} & -Y^{(2)}_{\bm{3}_0,1} \\
		-Y^{(2)}_{\bm{3}_0,2} & -Y^{(2)}_{\bm{3}_0,1} & 2Y^{(2)}_{\bm{3}_0,3}
	\end{pmatrix}\,.
\end{align}
the light neutrino Majorana mass matrix $m_{\nu}$ is given by the see-saw relation
\begin{equation}
M_\nu= -M_D^T M_N^{-1} M_D\,.
\end{equation}
Without loss of generality, the parameters $\alpha$, $\beta$, $\gamma$ and $g_1$ can be taken real, since their phases can be removed by field redefinition. Nevertheless the coupling $g_2$ is complex without imposing CP symmetry. As a consequence, all the lepton masses and mixing parameters only depend on six real parameters besides the complex modulus $\tau$. By scanning over the parameter space, if neutrino masses are normal hierarchy, we find that a good agreement between the model predictions and the experimental data can be achieved for the following choice of parameter values:
\begin{align}
\nonumber &\tau =0.103786+1.34097i\,,\quad \beta/\alpha= 2321.27\,,\quad \gamma/\alpha=798.326\,,\\
&g_2/g_1 = 13.8646 - 4.23100i\,,\quad \alpha v_d = 0.536619~ \text{MeV}\,,\quad g_1^2v^2_u/\Lambda= 5.84234~\text{meV}\,.
\end{align}
The lepton masses and mixing parameters at this point are determined to be:
\begin{align}
	\nonumber &\sin^2\theta_{12}=0.315878\,,\quad \sin^2\theta_{13}=0.021913\,,\quad \sin^2\theta_{23}=0.531526\,,\quad \delta_{CP}=1.13711\pi\,,\\
	\nonumber &
	\alpha_{21}= 1.03704\pi \,,~~\alpha_{31}= 1.15945\pi \,,~~ m_e/m_\mu=0.00480007 ,~~m_\mu/m_\tau=0.0566796\,,  \\
&m_1=6.47366~\text{meV}\,,\quad m_2 =10.7754~\text{meV}\,,\quad m_3 =49.9964~\text{meV}\,,\label{eq:nu-masses-mixpar-ist}
\end{align}
which are within the $3\sigma$ intervals of the experimental data without SK atmospheric data~\cite{Esteban:2020cvm}. The neutrino masses are of normal hierarchy type, and the neutrino mass sum is fixed to be
\begin{equation}
m_1+m_2+m_3=67.2454 ~\mathrm{meV}\,,
\end{equation}
which is well compatible with the latest result $\sum_i m_i < 120$ meV obtained from the combination of Planck TT, TE, EE+lowE+lensing+BAO data~\cite{Aghanim:2018eyx}. From the predicted values of mixing angles and neutrino masses in Eq.~\eqref{eq:nu-masses-mixpar-ist}, one can determine the effective electron anti-neutrino mass $m_{\beta}=\sqrt{\sum_i |U_{ei}|^2m^2_i}$ measured in $\beta$ decay experiments and the effective Majorana neutrino mass $m_{\beta\beta}=|\sum_iU^2_{ei}m_i|$ characterizing the amplitude of neutrinoless double decay,
\begin{equation}
m_\beta=10.8945~\text{meV} \,,\quad m_{\beta\beta}=2.63536\times 10^{-6}~\text{meV}\,.
\end{equation}
The quantity $m_{\beta}$ is much smaller than the KATRIN tritium beta decay upper limit $m_{\beta}<0.8$~eV at $90\%$ confidence level~\cite{Aker:2021gma,KATRIN:2019yun}. A strong cancellation in $m_{\beta\beta}$ occurs such that $m_{\beta\beta}$ almost vanishes. On the other hand, our model would be disfavoured if a positive signal is found.

This model can also accommodate inverted ordering neutrino masses in some region of parameter space. The agreement between predictions and data without SK~\cite{Esteban:2020cvm} is optimized by the following values of free parameters:
\begin{align}
	\nonumber &\tau =-0.497747+1.25636i\,,\quad \beta/\alpha= 0.056594\,,\quad \gamma/\alpha=0.0017391\,,\\
	&g_2/g_1 = -0.494173 + 0.23773 i\,,\quad \alpha v_d = 1227.75~ \text{MeV}\,,\quad g_1^2v^2_u/\Lambda= 538.424~\text{meV}\,.
\end{align}
The corresponding predictions for lepton masses, mixing angles, and CP violating phases as well as the effective neutrino masses $m_{\beta}$ and $m_{\beta\beta}$ are
\begin{align}
	\nonumber &\sin^2\theta_{12}=0.303001\,,\quad \sin^2\theta_{13}=0.0224254\,,\quad \sin^2\theta_{23}=0.520345\,,\\
	\nonumber &\delta_{CP}=1.77550\pi\,,\quad
	\alpha_{21}= 1.23377\pi \,,~~\alpha_{31}= 0.525454\pi \,,\\
\nonumber& m_e/m_\mu=0.00480000 ,~~m_\mu/m_\tau=0.0565946\,, \\
\nonumber&m_1=57.4753~\text{meV}\,, \quad m_2 =58.1172~\text{meV}\,,\quad m_3 =29.8200~\text{meV}\,,\\
	&m_\beta=57.1948~\text{meV} \,,\quad m_{\beta\beta}=28.1651\text{meV}\,.
\end{align}
The most stringent upper limit on $m_{\beta\beta}$ at present is provided by the KamLAND-Zen collaboration~\cite{KamLAND-Zen:2016pfg},
\begin{equation}
m_{\beta\beta}<(61-165)\; \text{meV} ~\mathrm{at}~ 90\%\;\mathrm{ C.L.} \,,
\end{equation}
where the uncertainty comes from the calculations of the numerical matrix elements. Future tonne-scale experiments such as nEXO~\cite{nEXO:2021ujk} and LEGEND~\cite{LEGEND:2021bnm} aim to probe the full region of the inverted neutrino mass ordering and a significant portion of the normal ordering can be explored. They intend to improve their sensitivity by a factor of approximately ten,
\begin{eqnarray}
m_{\beta\beta}<\left\{
\begin{array}{ll}
(4.7-20.3)\;\text{meV},&~~~~\text{nEXO}\,,\\
(9-21)\;\text{meV}, & ~~~~\text{LEGEND}-1000\,.
\end{array}
\right.
\end{eqnarray}
We see that our prediction of $m_{\beta\beta}$ for the inverted ordering is within the reach of nEXO and LEGEND sensitivities. Moreover, the neutrino mass sum is $m_1+m_2+m_3=145.413$ meV in this case, it is slightly larger than the aggressive upper bound of  $\sum_i m_i<120$meV although still compatible with the conservative bound $m_1+m_2+m_3<600$ meV extracted from the measurement of Planck lensing+BAO+$\theta_{\text{MC}}$~\cite{Aghanim:2018eyx}. Notice that the neutrino mass limits depend not only the data sets analyzed but also on the cosmological model. The limits also become weaker when one departs from the framework of $\Lambda$CDM plus neutrino mass to frameworks with more cosmological parameters.

\subsection{Model 2 based on finite modular group $GL(2,3)$}

In this section, we proceed to consider another finite modular group $GL(2,3)$ and model construction for lepton masses and mixing. The group $GL(2,3)$ is the Schur covering of $S_4$ of ``+'' type, and its representations are those of $S_4$ plus two doublets $\bm{2}',\bm{2}''$ and one quartet $\bm{4}$. We summarize the group theory of $GL(2,3)$ in Appendix~\ref{app:GL(2,3)_group}. It is remarkable that $GL(2,3)$ can produce interesting flavor structures which are unavailable in both $S_4$ and $S'_4$~\cite{Liu:2020akv,Novichkov:2020eep} which is the double cover of $S_4$.

Since the representation matrix $\rho(S^2)=\mathbb{1}_d$  for all irreps of $GL(2,3)$, the VVMFs transforming in the irreps of $GL(2,3)$ must be of even weight. From the table~\ref{tab:IrrepProfile} or the dimension formula of VVMF in Eq.~\eqref{eq:dim-formula}, we can find that there are three modular multiplets at weight 2 transforming as $\bm{2}$, $\bm{2'}$, $\bm{3}$ under $GL(2,3)$. They can be constructed in terms of the generalized hypergeometric series, as shown in section~\ref{sec:resultVVMF}. Using the general results of Eqs.~(\ref{eq:2dVVMF}, \ref{eq:3dVVMF}), we obtain
\begin{align}
\label{eq:wt2GL23VVMF}
&Y^{(2)}_{\bm{2}}(\tau)=\begin{pmatrix}
\eta^4(\tau)(\frac{K(\tau)}{1728})^{-\frac{1}{6}}~ {}_2F_1(-\frac{1}{6},\frac{1}{6};\frac{1}{2};K(\tau)) \\
8\sqrt{3}\eta^4(\tau)(\frac{K(\tau)}{1728})^{\frac{1}{3}}~ {}_2F_1(\frac{1}{3},\frac{2}{3};\frac{3}{2};K(\tau)) \\
	\end{pmatrix}\,,\\
&Y^{(2)}_{\bm{2}'}(\tau)=\begin{pmatrix}
	-2 \eta^4(\tau)(\frac{K(\tau)}{1728})^{\frac{5}{24}}~ {}_2F_1(\frac{5}{24},\frac{13}{24};\frac{5}{4};K(\tau)) \\
	\eta^4(\tau)(\frac{K(\tau)}{1728})^{-\frac{1}{24}}~ {}_2F_1(-\frac{1}{24},\frac{7}{24};\frac{3}{4};K(\tau)) \\
\end{pmatrix}\,,\\
&Y^{(2)}_{\bm{3}}(\tau)=\begin{pmatrix}
	\eta^4(\tau)(\frac{K(\tau)}{1728})^{-\frac{1}{6}}~ {}_3F_2(-\frac{1}{6},\frac{1}{6},\frac{1}{2};\frac{3}{4},\frac{1}{4};K(\tau)) \\
	-4\sqrt{2} \eta^4(\tau)(\frac{K(\tau)}{1728})^{\frac{1}{12}}~ {}_3F_2(\frac{1}{12},\frac{5}{12},\frac{3}{4};\frac{1}{2},\frac{5}{4};K(\tau)) \\
	-16\sqrt{2} \eta^4(\tau)(\frac{K(\tau)}{1728})^{\frac{7}{12}}~ {}_3F_2(\frac{7}{12},\frac{11}{12},\frac{5}{4};\frac{7}{4},\frac{3}{2};K(\tau))
\end{pmatrix}\,.
\end{align}
The $q$-expansions for these modular forms are given by
\begin{align}
\label{eq:q-wt2GL23VVMF}
\nonumber
&Y^{(2)}_{\bm{2}}(\tau)=\begin{pmatrix}
1 + 24 q + 24 q^2 + 96 q^3 + 24 q^4 + 144 q^5 + 96 q^6 + \dots \\
8\sqrt{3} q^{1/2} (1 + 4 q + 6 q^2 + 8 q^3 + 13 q^4 + 12 q^5 + 14 q^6 + \dots)\\
\end{pmatrix}\,,\\
\nonumber
&Y^{(2)}_{\bm{2}'}(\tau)=\begin{pmatrix}
-2 q^{3/8} (1 - 3 q + q^2 + 2 q^3 + 5 q^5 - 6 q^6 + \dots) \\
 q^{1/8} (1 -  q - 6 q^2 + 5 q^3 + 12 q^4 - 6 q^5 - 7 q^6 + \dots)\\
\end{pmatrix}\,,\\
&Y^{(2)}_{\bm{3}}(\tau)=\begin{pmatrix}
1 - 8 q + 24 q^2 - 32 q^3 + 24 q^4 - 48 q^5 + 96 q^6 + \dots \\
-4\sqrt{2}	q^{1/4} (1 + 6 q + 13 q^2 + 14 q^3 + 18 q^4 + 32 q^5 + 31 q^6 + \dots)\\
-16\sqrt{2}	q^{3/4} (1 + 2 q + 3 q^2 + 6 q^3 + 5 q^4 + 6 q^5 + 10 q^6 + \dots)\\
\end{pmatrix}\,,
\end{align}
Note that the components of $Y^{(2)}_{\bm{2}}(\tau)$ and $Y^{(2)}_{\bm{3}}(\tau)$ are scalar valued modular forms of $\Gamma(4)$. Based on the lowest modular forms $Y^{(2)}_{\bm{2}}$, $Y^{(2)}_{\bm{2}'}$ and $Y^{(2)}_{\bm{3}}$, one can construct the modular forms of higher weights by using the decomposition rules of $GL(2,3)$ listed in tables~\ref{tab:GL23_CG-1st} and~\ref{tab:GL23_CG-2nd}. At weight 4, there are totally six independent VVMFs given by
\begin{align}
\nonumber &Y^{(4)}_{\bm{1}}= (Y^{(2)}_{\bm{2}} Y^{(2)}_{\bm{2}})_{\bm{1}} = (Y^{(2)}_{\bm{2},1})^2+(Y^{(2)}_{\bm{2},2})^2 \,, \\
\nonumber &Y^{(4)}_{\bm{2}}= (Y^{(2)}_{\bm{2}} Y^{(2)}_{\bm{2}})_{\bm{2}} = \begin{pmatrix}
(Y^{(2)}_{\bm{2},2})^2-(Y^{(2)}_{\bm{2},1})^2 \\
2 Y^{(2)}_{\bm{2},1}Y^{(2)}_{\bm{2},2} \\
\end{pmatrix}\,,\\
\nonumber &Y^{(4)}_{\bm{2}'}= (Y^{(2)}_{\bm{2}'} Y^{(2)}_{\bm{3}})_{\bm{2}'}= \begin{pmatrix}
		\sqrt{2}Y^{(2)}_{\bm{2}',2}Y^{(2)}_{\bm{3},2}+Y^{(2)}_{\bm{2}',1}Y^{(2)}_{\bm{3},1} \\ 				\sqrt{2}Y^{(2)}_{\bm{2}',1}Y^{(2)}_{\bm{3},3}-Y^{(2)}_{\bm{2}',2}Y^{(2)}_{\bm{3},1} \\
\end{pmatrix}\,,\\
\nonumber &Y^{(4)}_{\bm{3}}= (Y^{(2)}_{\bm{2}} Y^{(2)}_{\bm{3}})_{\bm{3}}= \begin{pmatrix}
-2 Y^{(2)}_{\bm{2},1}Y^{(2)}_{\bm{3},1} \\
Y^{(2)}_{\bm{2},1}Y^{(2)}_{\bm{3},2} +\sqrt{3}Y^{(2)}_{\bm{2},2}Y^{(2)}_{\bm{3},3} \\ Y^{(2)}_{\bm{2},1}Y^{(2)}_{\bm{3},3} +\sqrt{3}Y^{(2)}_{\bm{2},2}Y^{(2)}_{\bm{3},2} \\
\end{pmatrix}\,,\\
\nonumber&Y^{(4)}_{\bm{3}'}=\frac{1}{\sqrt{2}}(Y^{(2)}_{\bm{2}'}Y^{(2)}_{\bm{2}'})_{\bm{3}'}=
 \begin{pmatrix}
\sqrt{2} Y^{(2)}_{\bm{2}',1}Y^{(2)}_{\bm{2}',2} \\
-(Y^{(2)}_{\bm{2}',1})^2 \\
(Y^{(2)}_{\bm{2}',2})^2 \\
\end{pmatrix}\,,\\
&Y^{(4)}_{\bm{4}}=(Y^{(2)}_{\bm{2}}Y^{(2)}_{\bm{2}'})_{\bm{4}}=
\begin{pmatrix}
	Y^{(2)}_{\bm{2},2}Y^{(2)}_{\bm{2}',2} \\
	-Y^{(2)}_{\bm{2},1}Y^{(2)}_{\bm{2}',1} \\
	Y^{(2)}_{\bm{2},2}Y^{(2)}_{\bm{2}',1}\\
	Y^{(2)}_{\bm{2},1}Y^{(2)}_{\bm{2}',2}\\
\end{pmatrix}\,.
\end{align}
Notice that $Y^{(4)}_{\bm{1}}=E_4, Y^{(4)}_{\bm{2}}=6D_2 Y^{(2)}_{\bm{2}}, Y^{(4)}_{\bm{2}'}=24D_2 Y^{(2)}_{\bm{2}'}$ and $Y^{(4)}_{\bm{3}}=12D_2 Y^{(2)}_{\bm{3}}$.

\subsubsection{Structure of the model}

In the following, we present a viable lepton model based on the finite modular group $GL(2, 3)$. The neutrino masses arise from the seesaw mechanism. Both the left-handed lepton doublets $L$ and right-handed neutrinos $N^c$ are organized into two triplet $\mathbf{3}$ of $GL(2,3)$ with the modular weights $k_{L}=2$ and $k_{N^c}=0$. The first two generations of the right-handed charged leptons $E_D^c\equiv(E_1^c, E_2^c)^T$ are assigned to a doublet $\bm{2''}$ with weight $k_{E_D^c}=2$, and the third right-handed charged lepton $E_3^c$ transform as $\mathbf{1}$ with weight $k_{E_3^c}=0$. The Higgs doublets $H_u$ and $H_d$ are assumed to be $GL(2, 3)$ trivial singlets of zero modular weight. Then the superpotentials for the charged lepton and neutrino masses take the form
\begin{align}
\nonumber \mathcal{W}_e &=  \alpha \left( E_D^c L Y^{(4)}_{\bm{2}'}\right)_{\bm{1}} H_d + \beta \left(E_D^c L Y^{(4)}_{\bm{4}}\right)_{\bm{1}} H_d + \gamma \left(E_3^c L Y^{(2)}_{\bm{3}}\right)_{\bm{1}} H_d\,, \\
\mathcal{W}_\nu &=  g_1 \left(N^c L Y^{(2)}_{\bm{2}}\right)_{\bm{1}} H_u +g_2 \left( N^c L Y^{(2)}_{\bm{3}}\right)_{\bm{1}} H_u+ \Lambda \left(N^c N^c \right)_{\bm{1}}\,.
\end{align}
With the  Clebsch-Gordon coefficients of $GL(2,3)$ in Appendix~\ref{app:GL(2,3)_group}, we can straightforwardly read out the charged lepton and neutrino mass matrices as follows,
\begin{align}
\label{eq:GL23lept1}
\nonumber&M_e=  \begin{pmatrix}
-\alpha Y^{(4)}_{\bm{2}',2}+\sqrt{2}\beta Y^{(4)}_{\bm{4},4} & \sqrt{3}\beta Y^{(4)}_{\bm{4},3} & \sqrt{2}\alpha Y^{(4)}_{\bm{2}',1}-\beta Y^{(4)}_{\bm{4},2}  \\
\alpha Y^{(4)}_{\bm{2}',1}+\sqrt{2}\beta Y^{(4)}_{\bm{4},2} & \sqrt{2}\alpha Y^{(4)}_{\bm{2}',2}+\beta Y^{(4)}_{\bm{4},4} & \sqrt{3}\beta Y^{(4)}_{\bm{4},1} \\
\gamma Y^{(2)}_{\bm{3},1} & \gamma Y^{(2)}_{\bm{3},3} & \gamma Y^{(2)}_{\bm{3},2}
\end{pmatrix}v_d  \,, \\
\nonumber&M_D=\begin{pmatrix}
-2 g_1 Y^{(2)}_{\bm{2},1}  & -g_2  Y^{(2)}_{\bm{3},3} & g_2 Y^{(2)}_{\bm{3},2} \\
g_2 Y^{(2)}_{\bm{3},3}  & \sqrt{3}g_1 Y^{(2)}_{\bm{2},2} & g_1 Y^{(2)}_{\bm{2},1}-g_2Y^{(2)}_{\bm{3},1} \\
-g_2  Y^{(2)}_{\bm{3},2} & g_1 Y^{(2)}_{\bm{2},1}+g_2Y^{(2)}_{\bm{3},1} & \sqrt{3}g_1 Y^{(2)}_{\bm{2},2}
\end{pmatrix}v_u\,,\\
&M_N= \Lambda \begin{pmatrix}
	1 ~& 0 ~& 0 \\
	0 ~& 0 ~& 1 \\
	0 ~& 1 ~& 0
	\end{pmatrix}\,.
\end{align}
We can set the parameters $\alpha$, $\gamma$ and $g_1$ to be real, since their phases can be absorbed into matter fields. The other two couplings $\beta$ and $g_2$ are complex, Through numerical search, we can find the best fit values of the parameters compatible with normal ordering data~\cite{Esteban:2020cvm} are given by:
\begin{align}
\nonumber &\tau =-0.499623+0.903453i\,,\quad \beta/\alpha= -2.90228 - 0.156823i\,,\quad \gamma/\alpha=0.000752696\,,\\
&g_2/g_1 =0.204168 - 0.909287i\,,\quad \alpha v_d = 327.660~ \text{MeV}\,,\quad g_1^2v^2_u/\Lambda=18.4597~\text{meV}\,.
\end{align}
The value of $\tau$ is very close to the left boundary of the fundamental domain. The predictions for the mass eigenvalues, mixing angles and CP violation phases are given by
\begin{align}
\nonumber &\sin^2\theta_{12}=0.303987\,,\quad \sin^2\theta_{13}=0.0222024\,,\quad \sin^2\theta_{23}=0.578472\,,\\ \nonumber &\delta_{CP}=1.59938\pi\,,
\quad \alpha_{21}= 1.44951\pi \,,\quad \alpha_{31}= 0.518406\pi\,,\\
\nonumber&m_e/m_\mu=0.00480015\,,~~~ m_\mu/m_\tau=0.0566693\,, \\
\nonumber& m_1=76.4654~\text{meV}\,,\quad m_2 =76.9491~\text{meV}\,,\quad  m_3 =91.4362~\text{meV}\,,\\
& m_\beta=76.9729~\text{meV} \,,\quad m_{\beta\beta}=53.2091~\text{meV}\,.
\end{align}	
All the experimental bounds in neutrino oscillation, tritium beta decays and neutrinoless double beta decay are satisfied. The neutrino mass sum is $m_1+m_2+m_2=235.45$ meV, this is compatible with the conservative bound $\sum_i m_i < 600$ meV from cosmology although larger than the most aggressive upper bound $\sum_i m_i < 120$ meV. This model can also fit the data of inverted neutrino mass spectrum. As an example, we give a point in the parameter space:
\begin{align}
\nonumber &\tau =-0.474696+0.880646i\,,\quad \beta/\alpha= -2.99553 - 0.191581i\,,\quad \gamma/\alpha=0.000992565\,,\\
&g_2/g_1 =0.813157 + 1.55181i\,,\quad \alpha v_d = 305.208~ \text{MeV}\,,\quad g_1^2v^2_u/\Lambda=13.0640~\text{meV}\,,
\end{align}
{The corresponding flavor observables are determined as:}
which leads to the following values for the flavor observables
\begin{align}
\nonumber &\sin^2\theta_{12}=0.304346\,,\quad \sin^2\theta_{13}=0.0224044\,,\quad \sin^2\theta_{23}=0.611139\,,\\ \nonumber &\delta_{CP}=1.58351\pi\,,
\quad \alpha_{21}= 0.745402\pi \,,\quad \alpha_{31}= 1.56947\pi\,,\\
\nonumber&m_e/m_\mu=0.00480415\,,~\quad~ m_\mu/m_\tau=0.0565167\,, \\
\nonumber& m_1=101.897~\text{meV}\,,\quad m_2 =102.260~\text{meV}\,, \quad m_3 =89.2362~\text{meV}\,,\\
& m_\beta=101.739~\text{meV} \,,\quad m_{\beta\beta}=54.1723~\text{meV}\,.
\end{align}
Similarly to previous case, all the current experimental bounds from neutrino oscillation, endpoint $\beta$ decay, neutrinoless double beta decay as well as conservative bound from cosmology are fulfilled.

\section{\label{sec:conclusion}Summary and outlook}

The modular flavor symmetry is an important theoretical advance to address the flavor puzzle of the SM, it is assumed that the Yukawa couplings are modular forms of the principal congruence subgroup $\Gamma(N)$ and the matter fields transform in representations of the homogeneous (inhomogeneous) finite modular group $\Gamma'_N$ ($\Gamma_N$). In the present work, we show that the requirement of modular forms of $\Gamma(N)$ is not a necessity. One can start from the vector-valued modular form which is a holomorphic function of $\tau$ and satisfies $Y(\gamma\tau)=(c\tau+d)^k\rho(\gamma)Y(\tau)$. Here $k$ is the modular weight, $\rho$ is an irrep of the modular group  $SL(2,\mathbb{Z})$ and it is assumed to have finite image. The vector-valued modular form $Y(\tau)$ is the scalar modular form of the kernel $\ker(\rho)$, and the image $\text{Im}(\rho)\cong \Gamma/\ker(\rho)$ plays the role of discrete finite flavor symmetry. The case of $\ker(\rho)=\Gamma(N)$ is considered in the original framework of modular flavor symmetry. The VVMFs can considerably extend the modular flavor symmetry, and we have more possible choices for the finite modular groups to construct modular invariant models. The general finite modular groups up to order 72 are listed in table~\ref{tab:NorSubgroupSL2Z}, they provide new opportunity for future study.

The theory of vector-valued modular form is reviewed in section~\ref{sec:VVMF}. The low dimensional vector-valued modular forms at the minimal weight can be determined by solving the modular linear differential equation. The one-dimensional vector-valued modular forms are powers of the Dedekind eta function. The two-dimensional and three-dimensional vector-valued modular forms can be expressed in terms of the generalized hypergeometric series. The space of vector-valued modular forms $\mathcal{M}(\rho)$ in representation $\rho$ is a module over the ring $\mathbb{C}[E_4,E_6]$, and it can be generated from a set of basis by multiplying polynomials of $E_4, E_6$.
Hence the modular multiplets in any representation $\rho$ have a finite number of $\tau$ dependent alignments, then only a finite number of models could be constructed with a finite modular groups, although there are infinite possible weight and representation assignments for the matter fields. Guided by the theory of vector-valued modular form, the construction of high weight modular multiplets is much easier, it is not necessary to  search for non-linear constraints as in the approach of tensor products.

We have provided examples of applying our formalism in section~\ref{sec:example}, and constructed two lepton
flavor models based on the finite modular groups $A_4\times Z_2$ and $GL(2,3)$. The neutrino masses are generated by the type I seesaw mechanism and no flavon field other than the modulus $\tau$ is used. The neutrino masses can be with either normal ordering or inverted ordering in both models, and the experimental data from neutrino oscillation, tritium beta decay, neutrinoless double beta decay and cosmology can be accommodated in certain parameter space. It is known that CP symmetry can be consistently combined with modular symmetrys~\cite{Novichkov:2019sqv,Baur:2019kwi}, and the modulus transforms as $\tau\rightarrow-\tau^{*}$ under the action of CP. 
The CP symmetry can constrain the phases of the coupling constants, and thus make the modular models more predictive. It is interesting to discuss the implications of CP symmetry in the context of modular flavor symmetry extended by vector-valued modular forms in future.

Although we have focused on the modular group $SL(2,\mathbb{Z})$, the modular invariance framework can be extended to more general Fuchsian group of genus zero, such as $\Gamma(2)$\cite{Li:2021buv} and $\Gamma_0(2)$~\cite{gottesman2020arithmetic}, even the symplectic modular group $Sp(2g,\mathbb{Z})$~\cite{Ding:2020zxw} and so on. The well-defined vector-valued modular forms exist for these group.
It would be interesting to explore the application of their vector-valued modular forms to the flavor puzzle. Finally we mention that the matter fields and Yukawa couplings transform as irreducible representations of $SL(2,\mathbb{Z})$ in top-down constructions, and the flavor symmetry group could be read from the image of the representations. We expect that the bottom-up analyses in view of vector-valued modular forms could correspond to some top-down realizations based
on ultraviolet-complete theories.

\subsection*{Acknowledgements}
This work is supported by the National Natural Science Foundation of China under Grant Nos 11975224, 11835013 and 11947301 and the Key Research Program of the Chinese Academy of Sciences under Grant NO. XDPB15.

\newpage
\section*{Appendix}

\setcounter{equation}{0}
\renewcommand{\theequation}{\thesection.\arabic{equation}}

\begin{appendix}

\section{\label{app:2-d irrep}All 2-d irreducible representations of $SL(2,\mathbb{Z})$ with finite image }

There are only four finite subgroups of $SU(2)$: the binary dihedral group $2D_{3}$, the binary tetrahedral group $T'$, the binary octahedral group $2O$ and the binary isosahedral group $A'_5$, which are homomorphic to $SL(2,\mathbb{Z})$~\cite{mason20082}. All the 2-d irreps of $SL(2,\mathbb{Z})$ with finite image are the direct product of the 1-d irreps of $SL(2,\mathbb{Z})$ with the 2-d irreps of the finite groups $2D_{3}$, $T'$, $2O$ and $A'_5$~\cite{mason20082}.  It is found that there are totally 54 inequivalent irreps with finite image, and they are listed in table~\ref{tab:2-d irrep} in terms of the parameters $r_1$ and $r_2$~\cite{mason20082}.

\begin{table}[h!]
\centering
\begin{tabular}{|c|c|c|c|}
\hline  \hline
\multicolumn{4}{|c|}{Binary dihedral type} \\  \hline
 & & &  \\[-0.18in] \hline	

$r_1$ ~&~ $r_2$ ~&~ $N$ ~&~ $k_0$ \\ \hline

$3/4$& $1/4$ & $4$ & $5$ \\ \hline 		

$5/6$& $1/3$ & $6$ & $6$ \\ \hline

$11/12$& $5/12$ & $12$ & $7$ \\ \hline

$0$& $1/2$ & $2$ & $2$ \\ \hline

$1/12$& $7/12$ & $12$ & $3$ \\ \hline

$1/6$& $2/3$ & $6$ & $4$ \\ \hline \hline
\end{tabular} \\
\begin{tabular}{|c|c|c|c|}
	\hline  \hline
	\multicolumn{4}{|c|}{Binary tetrahedral type} \\  \hline
	& & &  \\[-0.18in] \hline	
	
	$r_1$ ~&~ $r_2$ ~&~ $N$ ~&~ $k_0$ \\ \hline
	
	$2/3$& $1/3$ & $3$ & $5$ \\ \hline 		
	
	$3/4$& $5/12$ & $12$ & $6$ \\ \hline
	
	$5/6$& $1/2$ & $6$ & $7$ \\ \hline
	
	$11/12$& $7/12$ & $12$ & $8$ \\ \hline
	
	$0$& $2/3$ & $3$ & $3$ \\ \hline
	
	$1/12$& $3/4$ & $12$ & $4$ \\ \hline

	$1/6$& $5/6$ & $6$ & $5$ \\ \hline
	
	$1/4$& $11/12$ & $12$ & $6$ \\ \hline
	
	$1/3$& $0$ & $3$ & $1$ \\ \hline
	
	$5/12$& $1/12$ & $12$ & $2$ \\ \hline
	
	$1/2$& $1/6$ & $6$ & $3$ \\ \hline
	
	$7/12$& $1/4$ & $12$ & $4$ \\ \hline 			
\end{tabular}~~~
\begin{tabular}{|c|c|c|c|}
	\hline  \hline
	\multicolumn{4}{|c|}{Binary octahedral type} \\  \hline
	& & &  \\[-0.18in] \hline	
	
	$r_1$ ~&~ $r_2$ ~&~ $N$ ~&~ $k_0$ \\ \hline
	
	$7/8$& $1/8$ & $8$ & $5$ \\ \hline 		
	
	$23/24$& $5/24$ & $24$ & $6$ \\ \hline
	
	$1/24$& $7/24$ & $24$ & $1$ \\ \hline
	
	$1/8$& $3/8$ & $8$ & $2$ \\ \hline
	
	$5/24$& $11/24$ & $24$ & $3$ \\ \hline
	
	$7/24$& $13/24$ & $24$ & $4$ \\ \hline
	
	$3/8$& $5/8$ & $8$ & $5$ \\ \hline
	
	$11/24$& $17/24$ & $24$ & $6$ \\ \hline
	
	$13/24$& $19/24$ & $24$ & $7$ \\ \hline
	
	$5/8$& $7/8$ & $8$ & $8$ \\ \hline
	
	$17/24$& $23/24$ & $24$ & $9$ \\ \hline
	
	$19/24$& $1/24$ & $24$ & $4$ \\ \hline 			
\end{tabular} \\
\begin{tabular}{|c|c|c|c|}
	\hline  \hline
	\multicolumn{4}{|c|}{Binary Icosahedral type I} \\  \hline
	& & &  \\[-0.18in] \hline	
	
	$r_1$ ~&~ $r_2$ ~&~ $N$ ~&~ $k_0$ \\ \hline
	
	$4/5$& $1/5$ & $5$ & $5$ \\ \hline 		
	
	$53/60$& $17/60$ & $60$ & $6$ \\ \hline
	
	$29/30$& $11/30$ & $30$ & $7$ \\ \hline
	
	$9/20$& $1/20$ & $20$ & $2$ \\ \hline
	
	$2/15$& $8/15$ & $15$ & $3$ \\ \hline
	
	$13/60$& $37/60$ & $60$ & $4$ \\ \hline
	
	$3/10$& $7/10$ & $10$ & $5$ \\ \hline
	
	$23/60$& $47/60$ & $60$ & $6$ \\ \hline
	
	$7/15$& $13/15$ & $15$ & $7$ \\ \hline
	
	$11/20$& $19/20$ & $20$ & $8$ \\ \hline
	
	$19/30$& $1/30$ & $30$ & $3$ \\ \hline
	
	$43/60$& $7/60$ & $60$ & $4$ \\ \hline 			
\end{tabular}~~~
\begin{tabular}{|c|c|c|c|}
	\hline  \hline
	\multicolumn{4}{|c|}{Binary Icosahedral type II} \\  \hline
	& & &  \\[-0.18in] \hline	
	
	$r_1$ ~&~ $r_2$ ~&~ $N$ ~&~ $k_0$ \\ \hline
	
	$3/5$& $2/5$ & $5$ & $5$ \\ \hline 		
	
	$41/60$& $29/60$ & $60$ & $6$ \\ \hline
	
	$23/30$& $17/30$ & $30$ & $7$ \\ \hline
	
	$17/20$& $13/20$ & $20$ & $8$ \\ \hline
	
	$14/15$& $11/15$ & $15$ & $9$ \\ \hline
	
	$1/60$& $49/60$ & $60$ & $4$ \\ \hline
	
	$1/10$& $9/10$ & $10$ & $5$ \\ \hline
	
	$11/60$& $59/60$ & $60$ & $4$ \\ \hline
	
	$4/15$& $1/15$ & $15$ & $1$ \\ \hline
	
	$7/20$& $3/20$ & $20$ & $2$ \\ \hline
	
	$13/30$& $7/30$ & $30$ & $3$ \\ \hline
	
	$31/60$& $19/60$ & $60$ & $4$ \\ \hline 			
\end{tabular}
\caption{\label{tab:2-d irrep}Summary of the 54 irreducible 2-d representations of $SL(2,\mathbb{Z})$ with finite image, where $N$ denotes the order of $\rho(T)$. The unordered pair $(r_1,r_2)$ uniquely determines the 2-d irreps up to a similar transformation, and the representation matrices $\rho(S)$ and $\rho(T)$ are given in Eqs.~(\ref{eq:DiagT-2d},\ref{eq:rhoS-2d}).}
\end{table}

\section{\label{app:VVMF-modules-A4xZ2}VVMF modules of the finite modular group $A_4\times Z_2$}

Traditionally the higher weight modular multiplets are built from the tensor product of the lower weight modular multiplets. For instance, at weight 6, we have the following six linearly independent modular multiplets of $A_4\times Z_2$,
\begin{align}
\nonumber &Y^{(6)}_{\bm{1}_0}= (Y^{(2)}_{\bm{3}_0} Y^{(4)}_{\bm{3}_0})_{\bm{1}_0}=(Y^{(2)}_{\bm{3}_0,1})^3+(Y^{(2)}_{\bm{3}_0,2})^3+(Y^{(2)}_{\bm{3}_0,3})^3-3Y^{(2)}_{\bm{3}_0,1}Y^{(2)}_{\bm{3}_0,2}Y^{(2)}_{\bm{3}_0,3} \,, \\
\nonumber &Y^{(6)}_{\bm{1}_1}= Y^{(2)}_{\bm{1}''_1}Y^{(4)}_{\bm{1}'}=(Y^{(2)}_{\bm{3}_0,3})^2Y^{(2)}_{\bm{1}''_1}+2Y^{(2)}_{\bm{3}_0,1}Y^{(2)}_{\bm{3}_0,2}Y^{(2)}_{\bm{1}''_1} \,, \\
\nonumber&Y^{(6)}_{\bm{1}''_1}= Y^{(2)}_{\bm{1}''_1} Y^{(4)}_{\bm{1}}=(Y^{(2)}_{\bm{3}_0,1})^2Y^{(2)}_{\bm{1}''_1}+2Y^{(2)}_{\bm{3}_0,2}Y^{(2)}_{\bm{3}_0,3}Y^{(2)}_{\bm{1}''_1} \,, \\
\nonumber &Y^{(6)}_{\bm{3}_0I}=Y^{(2)}_{\bm{3}_0} Y^{(4)}_{\bm{1}_0}=\begin{pmatrix}
(Y^{(2)}_{\bm{3}_0,1})^3+2Y^{(2)}_{\bm{3}_0,1}Y^{(2)}_{\bm{3}_0,2}Y^{(2)}_{\bm{3}_0,3} \\
(Y^{(2)}_{\bm{3}_0,1})^2Y^{(2)}_{\bm{3}_0,2}+2(Y^{(2)}_{\bm{3}_0,2})^2Y^{(2)}_{\bm{3}_0,3} \\
(Y^{(2)}_{\bm{3}_0,1})^2Y^{(2)}_{\bm{3}_0,3}+2(Y^{(2)}_{\bm{3}_0,3})^2Y^{(2)}_{\bm{3}_0,2}
\end{pmatrix}\,,\\
\nonumber &Y^{(6)}_{\bm{3}_0II}=Y^{(2)}_{\bm{3}_0} Y^{(4)}_{\bm{1}'_0}= \begin{pmatrix}
(Y^{(2)}_{\bm{3}_0,3})^3+2Y^{(2)}_{\bm{3}_0,1}Y^{(2)}_{\bm{3}_0,2}Y^{(2)}_{\bm{3}_0,3} \\
(Y^{(2)}_{\bm{3}_0,3})^2Y^{(2)}_{\bm{3}_0,1}+2(Y^{(2)}_{\bm{3}_0,1})^2Y^{(2)}_{\bm{3}_0,2} \\
(Y^{(2)}_{\bm{3}_0,3})^2Y^{(2)}_{\bm{3}_0,2}+2(Y^{(2)}_{\bm{3}_0,2})^2Y^{(2)}_{\bm{3}_0,1}
\end{pmatrix}\,,\\
&Y^{(6)}_{\bm{3}_1} = Y^{(2)}_{\bm{1}''_1} Y^{(4)}_{\bm{3}_0} =\begin{pmatrix}
Y^{(2)}_{\bm{3}_0,3}(Y^{(2)}_{\bm{1}''_1})^2 \\
Y^{(2)}_{\bm{3}_0,1}(Y^{(2)}_{\bm{1}''_1})^2\\
Y^{(2)}_{\bm{3}_0,2}(Y^{(2)}_{\bm{1}''_1})^2 \\
\end{pmatrix}\,,\label{eq:wt6MF-A4xZ2}
\end{align}
where $Y^{(6)}_{\bm{3}_0I}$ and $Y^{(6)}_{\bm{3}_0II}$ stand for the two independent weight 6 VVMFs transforming as $\bm{3}_0$ under $A_4\times Z_2$. Other higher weight VVMFs can also be obtained in a similar way. The theory of VVMF gives us new insight to the structure of the space of the modular multiplets. As shown in section~\ref{sec:VVMF}, all the modular multiplets  $\mathcal{M}(\rho) = \bigoplus^{\infty}_{k=0}\mathcal{M}_k(\rho)$ in a representation $\rho$ is a graded module over the ring $\mathcal{M}(\mathbf{1})=\mathbb{C}[E_4,E_6]$, and a set of basis of the module $\mathcal{M}(\rho)$ is sufficient to generate the whole space of modular forms. Hence the VVMFs $ \mathcal{M}(A_4\times Z_2)$ in the representations of the finite modular group $A_4\times Z_2$ can be organized by representations and it comprises eight VVMF modules
\begin{equation}
\mathcal{M}(A_4\times Z_2)=\mathcal{M}(\mathbf{1}_0)\oplus\mathcal{M}(\mathbf{1}_1)\oplus\mathcal{M}(\mathbf{1}'_0)\oplus \mathcal{M}(\mathbf{1}'_1) \oplus \mathcal{M}(\mathbf{1}''_0) \oplus \mathcal{M}(\mathbf{1}''_1) \oplus \mathcal{M}(\mathbf{3}_0) \oplus \mathcal{M}(\mathbf{3}_1)\,.
\end{equation}
From table~\ref{tab:IrrepProfile} and the representation matrix of the modular generator $T$, we can read off the minimal weight and the basis of each module, and the basis can be generated from the minimal weight VVMF by using the modular differential operator $D^n_k$ for the triplet modules $\mathcal{M}(\mathbf{3}_0)$ and $\mathcal{M}(\mathbf{3}_1)$. Namely, we have
\begin{align}
\nonumber&\mathcal{M}(\mathbf{1}_0)=\langle 1 \rangle\,,\\
\nonumber&\mathcal{M}(\mathbf{1}_1)=\langle \widetilde{Y}^{(6)}_{\mathbf{1}_1} \rangle\,,\\
\nonumber&\mathcal{M}(\mathbf{1}'_0)=\langle \widetilde{Y}^{(4)}_{\mathbf{1}'_0} \rangle\,,\\
\nonumber&\mathcal{M}(\mathbf{1}'_1)=\langle \widetilde{Y}^{(10)}_{\mathbf{1}'_1} \rangle\,,\\
\nonumber&\mathcal{M}(\mathbf{1}''_0)=\langle \widetilde{Y}^{(8)}_{\mathbf{1}''_0} \rangle\,,\\
\nonumber&\mathcal{M}(\mathbf{1}''_1)=\langle \widetilde{Y}^{(2)}_{\mathbf{1}''_1} \rangle\,,\\
\nonumber&\mathcal{M}(\mathbf{3}_0)=\langle \widetilde{Y}^{(2)}_{\mathbf{3}_0}, D_2 \widetilde{Y}^{(2)}_{\mathbf{3}_0}, D_2^2 \widetilde{Y}^{(2)}_{\mathbf{3}_0} \rangle\,,\\
\label{eq:modules-A4xZ2} &\mathcal{M}(\mathbf{3}_1)=\langle \widetilde{Y}^{(4)}_{\mathbf{3}_1}, D_4 \widetilde{Y}^{(4)}_{\mathbf{3}_1}, D_4^2 \widetilde{Y}^{(4)}_{\mathbf{3}_1} \rangle\,.
\end{align}
Furthermore the minimal weight VVMFs can be extracted from the solutions of corresponding MLDEs shown in section~\ref{sec:resultVVMF}, i.e.
\begin{align}
\label{eq:wt6VVMF}
\nonumber&\widetilde{Y}^{(6)}_{\mathbf{1}_1}(\tau)= \eta^{12}(\tau)\,,~\widetilde{Y}^{(4)}_{\mathbf{1}'_0}(\tau)=\eta^{8}(\tau)\,,~\widetilde{Y}^{(10)}_{\mathbf{1}'_1}(\tau)=\eta^{20}(\tau)\,,\\
\nonumber&\widetilde{Y}^{(8)}_{\mathbf{1}''_0}(\tau)=\eta^{16}(\tau)\,,~~~~\widetilde{Y}^{(2)}_{\mathbf{1}''_1}(\tau)=\eta^{4}(\tau)\,,\\
\nonumber&\widetilde{Y}^{(2)}_{\mathbf{3}_0}(\tau)=\begin{pmatrix}
	\eta^4(\tau)(\frac{K(\tau)}{1728})^{-\frac{1}{6}}~ {}_3F_2(-\frac{1}{6},\frac{1}{6},\frac{1}{2};\frac{2}{3},\frac{1}{3};K(\tau)) \\
	-6 \eta^4(\tau)(\frac{K(\tau)}{1728})^{\frac{1}{6}}~ {}_3F_2(\frac{1}{6},\frac{1}{2},\frac{5}{6};\frac{2}{3},\frac{4}{3};K(\tau)) \\
	-18 \eta^4(\tau)(\frac{K(\tau)}{1728})^{\frac{1}{2}}~ {}_3F_2(\frac{1}{2},\frac{5}{6},\frac{7}{6};\frac{5}{3},\frac{4}{3};K(\tau))
\end{pmatrix}\,,\\
&\widetilde{Y}^{(4)}_{\mathbf{3}_1}(\tau)=\begin{pmatrix}
	-6\eta^8(\tau)(\frac{K(\tau)}{1728})^{\frac{1}{6}}~ {}_3F_2(\frac{1}{6},\frac{1}{2},\frac{5}{6};\frac{2}{3},\frac{4}{3};K(\tau)) \\
	-18 \eta^8(\tau)(\frac{K(\tau)}{1728})^{\frac{1}{2}}~ {}_3F_2(\frac{1}{2},\frac{5}{6},\frac{7}{6};\frac{5}{3},\frac{4}{3};K(\tau)) \\
	 \eta^8(\tau)(\frac{K(\tau)}{1728})^{-\frac{1}{6}}~ {}_3F_2(-\frac{1}{6},\frac{1}{6},\frac{1}{2};\frac{2}{3},\frac{1}{3};K(\tau))
\end{pmatrix}\,.
\end{align}
The linearly independent modular multiplets of $A_4\times Z_2$ at each weight can be straightforwardly obtained by multiplying the polynomial of $E_4,E_6$ with the bases of modules in Eq.~\eqref{eq:modules-A4xZ2}
\begin{align}
\nonumber&k=2: \widetilde{Y}^{(2)}_{\mathbf{3}_0}\,, \widetilde{Y}^{(2)}_{\mathbf{1}''_1}\,,\\
\nonumber&k=4: E_4\,, \widetilde{Y}^{(4)}_{\mathbf{1}'_0}\,,D_2\widetilde{Y}^{(2)}_{\mathbf{3}_0}\,,\widetilde{Y}^{(4)}_{\mathbf{3}_1}\,,\\
\nonumber&k=6: E_6\,, \widetilde{Y}^{(6)}_{\mathbf{1}_1}\,,E_4\widetilde{Y}^{(2)}_{\mathbf{1}''_1}\,,D^2_2\widetilde{Y}^{(2)}_{\mathbf{3}_0}\,,E_4\widetilde{Y}^{(2)}_{\mathbf{3}_0}\,,D_4\widetilde{Y}^{(4)}_{\mathbf{3}_1}\,,\\
\label{eq:MF-multiplets-A4xZ2}&k=8: E_4^2\,,E_4\widetilde{Y}^{(4)}_{\mathbf{1}'_0}\,,\widetilde{Y}^{(8)}_{\mathbf{1}''_0}\,, E_6\widetilde{Y}^{(2)}_{\mathbf{1}''_1}\,,E_6\widetilde{Y}^{(2)}_{\mathbf{3}_0}\,,E_4D_2\widetilde{Y}^{(2)}_{\mathbf{3}_0}\,,E_4\widetilde{Y}^{(4)}_{\mathbf{3}_1}\,,D_4^2\widetilde{Y}^{(4)}_{\mathbf{3}_1}\,,
\end{align}
Comparing the $q$-expressions, one can see that the modular multiplets in Eq.~\eqref{eq:MF-multiplets-A4xZ2} match those built from tensor products in Eqs.~(\ref{eq:wt4MF-A4xZ2}, \ref{eq:wt6MF-A4xZ2}),
\begin{align}
\nonumber&Y^{(4)}_{\mathbf{1}_0}=E_4\,,\quad Y^{(4)}_{\mathbf{1}'_0}=-12\widetilde{Y}^{(4)}_{\mathbf{1}'_0}\,,\quad Y^{(4)}_{\mathbf{3}_0}=-6D_2 \widetilde{Y}^{(2)}_{\mathbf{3}_0}\,,\quad Y^{(4)}_{\mathbf{3}_1}=\widetilde{Y}^{(4)}_{\mathbf{3}_1}\,, \\ \nonumber&Y^{(6)}_{\mathbf{1}_0}=E_6\,,\quad Y^{(6)}_{\mathbf{1}_1}=-12\widetilde{Y}^{(6)}_{\mathbf{1}_1}\,,\quad Y^{(6)}_{\mathbf{1}''_1}=E_4\widetilde{Y}^{(2)}_{\mathbf{1}''_1}\,,\quad Y^{(6)}_{\mathbf{3}_0I}=E_4\widetilde{Y}^{(2)}_{\mathbf{3}_0}\,, \\
&Y^{(6)}_{\mathbf{3}_0II}=2E_4\widetilde{Y}^{(2)}_{\mathbf{3}_0}-36D_2^2\widetilde{Y}^{(2)}_{\mathbf{3}_0}\,, \quad Y^{(6)}_{\mathbf{3}_1}=-6D_4\widetilde{Y}^{(4)}_{\mathbf{3}_1}\,.
\end{align}
Therefore the module structure of VVMFs makes the construction of modular multiplets much easier, and one doesn't need to search for the non-linear constraints which relate dependent multiplets coming from tensor products. Since multiplication of the polynomial of $E_4$, $E_6$ doesn't change the alignment, the modular multiplets in any given representation only have finite possible $\tau$ dependent alignments. These two approaches of constructing modular
forms are complementary. For example, the MLDEs for some minimal weight VVMFs in high dimensional representations are difficult to be solved analytically,
but they may be obtained from the tensor products of lower weight VVMFs in low dimensional representations.

\section{\label{app:finiteness}Finite number of the modular invariant models}

There are no restrictions on the representation and weight assignments for the matter fields, and the dimension of space of modualar forms generally increases with the modular weight. Therefore it seems that one can construct infinite possible models for neutrino masses and mixing based on a finite modular group. In this section, we show that only a finite number of models could be constructed because the modular multiplets $\mathcal{M}(\rho)$ in any representation $\rho$ have a few possible alignments, as explained in section~\ref{app:VVMF-modules-A4xZ2}.
In the following, we use a concrete example to explain the finiteness of modular invariant models. The neutrino masses are generated from the Weinberg operator, we assign the lepton doublets $L$ and right-handed charged leptons $E^c$ to triplets of $A_4\times Z_2$, both Higgs doublets $H_u$ and $H_d$ are trivial singlets and their modular weights are assumed to be vanishing without loss of generality,
\begin{align}
\label{eq:irrepAssign}
\rho_{E^c}=\mathbf{3}_1,~~ \rho_{L} = \mathbf{3}_0,~~\rho_{H_u}=\rho_{H_d}=\mathbf{1}_0\,.
\end{align}
The superpotentials for lepton masses can be written as:
\begin{align}
\nonumber \mathcal{W}_e &=  \alpha_1 \left( Y^{(k_e)}_{\bm{3}_1}(E^c L)_{\bm{3}_1,S} \right)_{\bm{1}_0}H_d + \alpha_2\left( Y^{(k_e)}_{\bm{3}_1} (E^c L)_{\bm{3}_1,A} \right)_{\bm{1}_0} H_d+\alpha_3 Y^{(k_e)}_{\bm{1}_1}(E^c L)_{\bm{1}_1} H_d  \\
\nonumber&~~+\alpha_4 Y^{(k_e)}_{\bm{1}''_1}(E^c L)_{\bm{1}'_1} H_d +\alpha_5 Y^{(k_e)}_{\bm{1}'_1}(E^c L)_{\bm{1}''_1} H_d \,,\\
\nonumber\mathcal{W}_\nu &=  \frac{g_1}{\Lambda} \left(Y^{(k_\nu)}_{\bm{3}_0} (L L)_{\bm{3}_0,S} \right)_{\bm{1}_0}H_u H_u+\frac{g_2}{\Lambda} Y^{(k_\nu)}_{\bm{1}_0} (L L)_{\bm{1}_0}H_u H_u\\
&~~+\frac{g_3}{\Lambda} Y^{(k_\nu)}_{\bm{1}''_0} (L L)_{\bm{1}'_0}H_u H_u+\frac{g_4}{\Lambda} Y^{(k_\nu)}_{\bm{1}'_0} (L L)_{\bm{1}''_0}H_u H_u\,. \label{eq:superpotential-app}
\end{align}
The modular form singlets can be absorbed into the free couplings $\alpha_{3,4,5}$ and $g_{2,3,4}$. It can occur that there are several independent triplet modular forms $Y^{(k_e)}_{\bm{3}_1}$ and $Y^{(k_\nu)}_{\bm{3}_0}$. Then all terms with
arbitrary coefficients should be included and the superpotential $\mathcal{W}_{e, \nu}$ contains more parameters than those explicitly indicated. Modular invariance of the superpotential requires the following relations for the weights
\begin{equation}
k_e=k_{L}+k_{E^c},~~~~k_{\nu}=2k_L\,,
\end{equation}
where $k_L$ and $k_{E^c}$ are the modular weights of $L$ and $E^c$ respectively. By choosing the values of $k_{L}$ and $k_{E^c}$, we can achieve any integer values for the modular weights $k_e$ and $k_{\nu}$. Thus it seems there are infinite possible models differing in the values of $k_{e,\nu}$, but different choices of $k_{e, \nu}$ can effectively lead to the same model so that one has a finite number of possibilities. The key point is that, the modular multiplets $Y^{(k_e)}_{\bm{3}_1}$ and $Y^{(k_\nu)}_{\bm{3}_0}$ for any values of $k_{e, \nu}$ can be expressed in terms of the three independent bases of the modules $\mathcal{M}(\mathbf{3}_0)$ and $\mathcal{M}(\mathbf{3}_1)$ over the ring $\mathbb{C}[E_4,E_6]$.  Namely, they can always be written as
\begin{align}
\label{eq:Yk}
\nonumber&Y^{(k_e)}_{\bm{3}_1}= f_1(\tau) Y^{(4)}_{\bm{3}_1} + f_2(\tau) D_4Y^{(4)}_{\bm{3}_1}+ f_3(\tau) D_4^2 Y^{(4)}_{\bm{3}_1}\,, \\
&Y^{(k_\nu)}_{\bm{3}_0}= h_1(\tau) Y^{(2)}_{\bm{3}_0} + h_2(\tau) D_2Y^{(2)}_{\bm{3}_0}+ h_3(\tau) D_2^2Y^{(2)}_{\bm{3}_0}\,,
\end{align}
where $f_1(\tau)$, $f_2(\tau)$ and $f_{3}(\tau)$ are polynomials of $E_4$, $E_6$, and they are scalar modular forms of $SL(2,\mathbb{Z})$ with weights $k_e-4$, $k_e-6$ and $k_e-8$ respectively. Analogously $h_1(\tau)$, $h_2(\tau)$ and $h_{3}(\tau)$ are modular forms of $SL(2,\mathbb{Z})$ with weights $k_\nu-2$, $k_\nu-4$ and $k_\nu-6$ respectively. The corresponding functions are vanishing if the modular forms at some of these weights don't exist. As a consequence, we see that the superpotential $\mathcal{W}_{e, \nu}$ has at most 15 independent couplings for $L\sim\mathbf{3}_0$ and $E^c\sim\mathbf{3}_1$. Even after considering other possible representation assignments for the matter fields, one can only construct a finite number of modular invariant superpotential based on $A_4\times Z_2$. Hence the charged lepton and neutrino mass matrices can only take a finite number of possible forms as well. This conclusion also applies to all other finite modular groups.

We proceed to investigate the K\"ahler potential before closing this section. The K\"ahler potential of the lepton doublets $L$ is considered for illustration, and its most general form is given by
\begin{eqnarray}
\label{eq:Kahlerl}
\mathcal{K}_L=
(-i\tau+i\bar{\tau})^{-k_L} (L^\dagger L)_{\bm{1}}
+ \sum_{k_Y,\mathbf{r}_1,\mathbf{r}_2} (-i\tau+i\bar{\tau})^{-k_{L}+k_1} (L^\dagger Y^{(k_Y) \dagger}_{\mathbf{r}_1}L Y^{(k_Y)}_{\mathbf{r}_2})_{\mathbf{1}_0}+\mathrm{h.c.}\,,
\end{eqnarray}
where we suppress all coupling constants and each operator could have several independent $A_4\times Z_2$ contractions. Because the modular multiplets in any representation of $A_4\times Z_2$ have a few alignments, it is sufficient to only consider the modular multiplets $Y^{(k_Y)}_{\mathbf{r}_1}$ and $Y^{(k_Y)}_{\mathbf{r}_2}$ which are the bases of the modules of $A_4\times Z_2$ shown in Eq.~\eqref{eq:modules-A4xZ2}. Therefore the modular symmetry constrains the K\"ahler potential to depend on a finite (not infinite) number of parameters. The presence of the additional terms besides the minimal K\"ahler potential reduces the predictive power of modular flavor symmetry~\cite{Chen:2019ewa}.

\section{\label{app:GL(2,3)_group}$GL(2,3)$ group }

The group $GL(2,3)$ consists of $2\times2$ invertible matrices whose elements are added and multiplied as integers modulo 3. It is the Schur cover of the permutation group $S_4$ of ``+'' type. The group $GL(2,3)$ can be generated by the modular generators $S$ and $T$ satisfying the relations:
\begin{equation}
S^2=(ST)^3=T^8=ST^4ST^{-4}=1\,.
\end{equation}
It has 48 elements and the group ID is $[48,29]$ in \texttt{GAP}~\cite{GAP}. The element $T^4$ commutes with all group elements, and the cyclic group $Z^{T^4}_2\equiv\left\{1, T^4\right\}$ is the center, and the quotient group $GL(2,3)/Z^{T^4}_2$ is isomorphic to $S_4$.

In addition to the representations of $S_4$: $\bm{1}$, $\bm{1}'$, $\bm{2}$, $\bm{3}$ and $\bm{3}'$, $GL(2, 3)$ possesses two doublet representations $\bm{2}',\bm{2}''$ and one quartet representation $\bm{4}$. The explicit forms of the representation matrices $\rho(S)$ and $\rho(T)$ in each of the irreducible representations are given in the following,
\begin{eqnarray}
\nonumber&&\hskip-0.4in \mathbf{1}:~S=1\,,~~T=1\,, \\[+0.1in]
\nonumber&&\hskip-0.4in \mathbf{1}':~S=-1\,,~~T=-1\,, \\[+0.1in]
\nonumber&&\hskip-0.4in \mathbf{2}:~S=\dfrac{1}{2}\left(
	\begin{array}{cc}
		-1 & -\sqrt{3} \\
		-\sqrt{3}& 1 \\
	\end{array}
	\right)\,,~~T= \left(
	\begin{array}{cc}
		1 & 0 \\
		0 & -1 \\
	\end{array}
	\right)\,,\\[+0.1in]
\nonumber&&\hskip-0.4in \mathbf{2}':~S= \dfrac{1}{\sqrt{2}}\left(
	\begin{array}{cc}
		1 & 1 \\
		1 & -1 \\
	\end{array}
	\right)\,,~~T=\left(
	\begin{array}{cc}
		e^{3 \pi i/4} & 0 \\
		0 & e^{\pi i /4} \\
	\end{array}
	\right)\,, \\[+0.1in]
\nonumber&&\hskip-0.4in \mathbf{2}'':~S=\dfrac{1}{\sqrt{2}}\left(
	\begin{array}{cc}
		-1 & 1 \\
		1 & 1 \\
	\end{array}
	\right)\,,~~T=\left(
	\begin{array}{cc}
		e^{-\pi i /4} & 0 \\
		0 & e^{-3 \pi i/4} \\
	\end{array}
	\right)\,, \\[+0.1in]
\nonumber&&\hskip-0.4in \mathbf{3}:~ S=\dfrac{1}{2}\left(
	\begin{array}{ccc}
		0 & \sqrt{2} & \sqrt{2} \\
		\sqrt{2} & -1 & 1 \\
		\sqrt{2} & 1 & -1 \\
	\end{array}
	\right)\,,~~ T=\left(
	\begin{array}{ccc}
		1 & 0 & 0 \\
		0 & i & 0 \\
		0 & 0 & -i \\
	\end{array}
	\right) \,, \\ [+0.1in]
\nonumber&&\hskip-0.4in \mathbf{3'}:~ S=-\dfrac{1}{2}\left(
	\begin{array}{ccc}
		0 & \sqrt{2} & \sqrt{2} \\
		\sqrt{2} & -1 & 1 \\
		\sqrt{2} & 1 & -1 \\
	\end{array}
	\right)\,,~~ T=-\left(
	\begin{array}{ccc}
		1 & 0 & 0 \\
		0 & i & 0 \\
		0 & 0 & -i \\
	\end{array}
	\right) \\  [+0.1in]
&&\hskip-0.4in \mathbf{4}:~ S=\dfrac{1}{2 \sqrt{2}}\left(
	\begin{array}{cccc}
		-1 & \sqrt{3} & 1 & \sqrt{3} \\
		\sqrt{3} & -1 & \sqrt{3} & 1 \\
		1 & \sqrt{3} & 1 & -\sqrt{3} \\
		\sqrt{3} & 1 & -\sqrt{3} & 1 \\
	\end{array}
	\right)\,, ~~ T=\left(
	\begin{array}{cccc}
		e^{-3 \pi i/4} & 0 & 0 & 0 \\
		0 &e^{3 \pi i/4} & 0 & 0 \\
		0 & 0 & e^{-\pi i/4} & 0 \\
		0 & 0 & 0 & e^{\pi i/4} \\
	\end{array}
	\right)\,.
\end{eqnarray}
Notice that the two 1-d irreps $\bm{1}$ and $\bm{1}'$ correspond to $\mathbf{1}_0$ and $\mathbf{1}_6$ of $SL(2,\mathbb{Z})$, these three 2-d irreps $\mathbf{2}$, $\mathbf{2}'$ and $\mathbf{2}''$ are the doublet representations $\mathbf{2}_{(0,\frac{1}{2})}$, $\mathbf{2}_{(\frac{3}{8},\frac{1}{8})}$ and $\mathbf{2}_{(\frac{7}{8},\frac{5}{8})}$ of $SL(2,\mathbb{Z})$, and the two 3-d irreps $\mathbf{3}$ and $\mathbf{3}'$ correspond to $\mathbf{3}_{(0,\frac{1}{4}, \frac{3}{4})}$ and $\mathbf{3}_{(\frac{1}{2}, \frac{3}{4}, \frac{1}{4})}$ of $SL(2,\mathbb{Z})$. We see the representation matrix $\rho(T^4)=1$ for the irreps $\bm{1}$, $\bm{1}'$, $\bm{2}$, $\bm{3}, \bm{3}'$ and $\rho(T^4)=-1$ for the irreps $\bm{2}'$, $\bm{2}''$, $\bm{4}$. Hence there is no way to distinguish the group $GL(2, 3)$ from $S_4$ when working with the irreps $\bm{1}$, $\bm{1}'$, $\bm{2}$, $\bm{3}$ and $\bm{3}'$. The 48 elements of $GL(2,3)$ can be divided into 8 conjugacy classes, and the character table is shown in table~\ref{tab:character}, it is obtain by taking trace of the explicit representation matrices.
\begin{table}[th!]
\begin{center}
\renewcommand{\tabcolsep}{2.8mm}
\renewcommand{\arraystretch}{1.3}
\begin{tabular}{|c|c|c|c|c|c|c|c|c|c|c|c|c|c|c|c|c|c|c|c|c|c|c|}\hline\hline
\text{Classes} & $1C_1$ & $1C_2$ & $6C_8$ & $6C'_8$ & $6C_4$ & $8C_6$ & $8C_3$ & $12C'_2$  \\ \hline
Sample element & $1$ & $T^4$ & $T$ & $ST^2$ & $(T^2S)^2$ & $TST^3S$ & $TST^4$ & $TST^3$ \\ \hline
Order of class    & 1  & 1  & 6  & 6 & 6  & 8  & 8  & 12\\\hline
Order of element &  1  & 2  & 8  & 8  & 4 & 6  & 3  & 2 \\\hline
$\mathbf{1}$ & $1$ & $1$ & $1$ & $1$ & $1$ & $1$ & $1$ & $1$\\ \hline
$\mathbf{1}'$ & $1$ & $1$ & $-1$ & $-1$ & $1$ & $1$ & $1$ & $-1$\\ \hline
$\mathbf{2}$ & $2$ & $2$ & $0$ & $0$ & $2$ & $-1$ & $-1$ & $0$\\ \hline
$\mathbf{2}'$ & $2$ & $-2$ & $\sqrt{2}\, i$ & $-\sqrt{2}\, i$ & $0$ & $-1$ \
& $1$ & $0$\\ \hline
$\mathbf{2}''$ & $2$ & $-2$ & $-\sqrt{2}\,i$ & $\sqrt{2}\, i$ & $0$ & $-1$ \
& $1$ & $0$\\ \hline
$\mathbf{3}$ & $3$ & $3$ & $1$ & $1$ & $-1$ & $0$ & $0$ & $-1$\\ \hline
$\mathbf{3}'$ & $3$ & $3$ & $-1$ & $-1$ & $-1$ & $0$ & $0$ & $1$\\ \hline
$\mathbf{4}$ & $4$ & $-4$ & $0$ & $0$ & $0$ & $1$ & $-1$ & $0$\\ \hline\hline
\end{tabular}
\caption{\label{tab:character} The character table of the group $GL(2,3)$.}
\end{center}
\end{table}

We present the decompositions of tensor products of different $GL(2,3)$ irreps and the corresponding Clebsch-Gordan coefficients in our basis. We use $\alpha_i$ to indicate the elements of the first representation of the product and $\beta_i$ to indicate those of the second representation.
The results are collected in table~\ref{tab:GL23_CG-1st} and table~\ref{tab:GL23_CG-2nd}.

\begin{table}[ht!]
\centering
\resizebox{1.0\textwidth}{!}{
\begin{tabular}{|c|c|c|c|c|c|c|c|c|c|c|c|}\hline\hline
\multicolumn{3}{|c}{~~~~~~$\bm{1}' \otimes \bm{2} = \bm{2}$~~~~~~} & \multicolumn{3}{|c}{$\bm{1}' \otimes \bm{2}' = \bm{2}''$} & \multicolumn{3}{|c}{~~~~~~$\bm{1}' \otimes \bm{2}''
	=\bm{2}'$~~~~~~}&\multicolumn{3}{|c|}{$\bm{1}' \otimes \bm{4} = \bm{4}$}  \\ \hline
\multicolumn{3}{|c}{$\mathbf{2} \sim \begin{pmatrix}
		\alpha \beta_2 \\
		-\alpha \beta_1 \\
\end{pmatrix} $} &
\multicolumn{3}{|c}{$\mathbf{2}'' \sim \begin{pmatrix}
		-\alpha \beta_1 \\
		\alpha \beta_2 \\
\end{pmatrix}$} &
\multicolumn{3}{|c}{$\mathbf{2}' \sim \begin{pmatrix}
		-\alpha \beta_1 \\
		\alpha \beta_2 \\
	\end{pmatrix}$} &
\multicolumn{3}{|c|}{$\mathbf{4} \sim \begin{pmatrix}
	\alpha \beta_4 \\
	\alpha \beta_3 \\
	-\alpha \beta_2 \\
	-\alpha \beta_1 \\
\end{pmatrix}$}  \\ \hline\hline
			
\multicolumn{6}{|c}{~~~~~~~~~~~~~~~~$\mathbf{1} \otimes \mathbf{3} = \mathbf{1}' \otimes \mathbf{3}' = \mathbf{3}$~~~~~~~~~~~~~~~} & \multicolumn{6}{|c|}{$\mathbf{1} \otimes \mathbf{3}' = \mathbf{1}' \otimes \mathbf{3} = \mathbf{3}'$} \\ \hline
\multicolumn{6}{|c}{$\mathbf{3}\sim\begin{pmatrix}
	\alpha\beta_1 \\
	\alpha\beta_2 \\
	\alpha\beta_3 \\
\end{pmatrix} $}&
\multicolumn{6}{|c|}{$ \mathbf{3'}\sim\begin{pmatrix}
   \alpha\beta_1 \\
	\alpha\beta_2 \\
	\alpha\beta_3 \\
	\end{pmatrix} $} \\ \hline\hline
			
\multicolumn{4}{|c}{$\mathbf{2} \otimes \mathbf{2} = \mathbf{1_s} \oplus \mathbf{1'_a} \oplus \mathbf{2_s}$} & \multicolumn{4}{|c}{$\mathbf{2} \otimes \mathbf{2}' = \mathbf{4}$} & \multicolumn{4}{|c|}{$\mathbf{2} \otimes \mathbf{2}'' = \mathbf{4}$}   \\ \hline
\multicolumn{4}{|c}{  $ \begin{array}{l}
\mathbf{1_s}\sim \alpha_1 \beta_1+\alpha_2 \beta_2 \\
\mathbf{1'_a}\sim \alpha_2 \beta_1-\alpha_1 \beta_2 \\
\mathbf{2_s}\sim\begin{pmatrix}
	\alpha_2 \beta_2-\alpha_1 \beta_1 \\
	\alpha_2 \beta_1+\alpha_1 \beta_2 \\
\end{pmatrix} \\
\end{array} $ } &
\multicolumn{4}{|c}{ $\mathbf{4} \sim \begin{pmatrix}
		\alpha_2 \beta_2 \\
		-\alpha_1 \beta_1 \\
		\alpha_2 \beta_1 \\
		\alpha_1 \beta_2 \\
	\end{pmatrix}$} &
\multicolumn{4}{|c|}{ $\mathbf{4} \sim \begin{pmatrix}
		-\alpha_1 \beta_2 \\
		\alpha_2 \beta_1 \\
		\alpha_1 \beta_1 \\
		\alpha_2 \beta_2 \\
	\end{pmatrix} $}  \\ \hline\hline
			
\multicolumn{4}{|c}{$\mathbf{2'} \otimes \mathbf{2'} = \mathbf{1'_a} \oplus \mathbf{3'_s} $} & \multicolumn{4}{|c}{$\mathbf{2'} \otimes \mathbf{2''} = \mathbf{1} \oplus \mathbf{3}$} & \multicolumn{4}{|c|}{$\mathbf{2''} \otimes \mathbf{2''} = \mathbf{1'_a} \oplus \mathbf{3'_s}$}  \\ \hline
\multicolumn{4}{|c}{  $\begin{array}{l}
		\mathbf{1'_a}\sim \alpha_2 \beta_1-\alpha_1 \beta_2 \\
		\mathbf{3'_s}\sim\begin{pmatrix}
			\alpha_2 \beta_1+\alpha_1 \beta_2 \\
			-\sqrt{2} \alpha_1 \beta_1 \\
			\sqrt{2} \alpha_2 \beta_2 \\
		\end{pmatrix}\\
	\end{array}$ } &
\multicolumn{4}{|c}{ $ \begin{array}{l}
		\mathbf{1}\sim \alpha_2 \beta_1+\alpha_1 \beta_2 \\
		\mathbf{3}\sim\begin{pmatrix}
			\alpha_1 \beta_2-\alpha_2 \beta_1 \\
			\sqrt{2} \alpha_1 \beta_1 \\
			\sqrt{2} \alpha_2 \beta_2 \\
		\end{pmatrix}\\
	\end{array} $} &
\multicolumn{4}{|c|}{ $\begin{array}{l}
		\mathbf{1'_a}\sim \alpha_2 \beta_1-\alpha_1 \beta_2 \\
		\mathbf{3'_s}\sim\begin{pmatrix}
			\alpha_2 \beta_1+\alpha_1 \beta_2 \\
			\sqrt{2} \alpha_1 \beta_1 \\
			-\sqrt{2} \alpha_2 \beta_2 \\
		\end{pmatrix}\\
	\end{array} $}  \\ \hline\hline

\multicolumn{4}{|c}{$\mathbf{2} \otimes \mathbf{3} = \mathbf{3} \oplus \mathbf{3'}$} & \multicolumn{4}{|c}{$\mathbf{2} \otimes \mathbf{3'} = \mathbf{3} \oplus \mathbf{3'}$} & \multicolumn{4}{|c|}{$\mathbf{2'} \otimes \mathbf{3} = \mathbf{2'} \oplus \mathbf{4}$}  \\ \hline
\multicolumn{4}{|c}{  $ \begin{array}{l}
		\mathbf{3} \sim \begin{pmatrix}
			-2 \alpha_1 \beta_1 \\
			\alpha_1 \beta_2+\sqrt{3} \alpha_2 \beta_3 \\
			\sqrt{3} \alpha_2 \beta_2+\alpha_1 \beta_3 \\
		\end{pmatrix} \\
		\mathbf{3'} \sim \begin{pmatrix}
			-2 \alpha_2 \beta_1 \\
			\alpha_2 \beta_2-\sqrt{3} \alpha_1 \beta_3 \\
			\alpha_2 \beta_3-\sqrt{3} \alpha_1 \beta_2 \\
		\end{pmatrix}\\
	\end{array} $ } &
\multicolumn{4}{|c}{ $ \begin{array}{l}
		\mathbf{3} \sim \begin{pmatrix}
			-2 \alpha_2 \beta_1 \\
			\alpha_2 \beta_2-\sqrt{3} \alpha_1 \beta_3 \\
			\alpha_2 \beta_3-\sqrt{3} \alpha_1 \beta_2 \\
		\end{pmatrix} \\
		\mathbf{3'} \sim \begin{pmatrix}
			-2 \alpha_1 \beta_1 \\
			\alpha_1 \beta_2+\sqrt{3} \alpha_2 \beta_3 \\
			\sqrt{3} \alpha_2 \beta_2+\alpha_1 \beta_3 \\
		\end{pmatrix}\\
	\end{array} $} &
\multicolumn{4}{|c|}{ $\begin{array}{l}
		\mathbf{2'} \sim \begin{pmatrix}
			\alpha_1 \beta_1+\sqrt{2} \alpha_2 \beta_2 \\
			\sqrt{2} \alpha_1 \beta_3-\alpha_2 \beta_1 \\
		\end{pmatrix} \\
		\mathbf{4} \sim \begin{pmatrix}
			\sqrt{3} \alpha_1 \beta_2 \\
			\sqrt{2} \alpha_1 \beta_1-\alpha_2 \beta_2 \\
			\sqrt{3} \alpha_2 \beta_3 \\
			\sqrt{2} \alpha_2 \beta_1+\alpha_1 \beta_3 \\
		\end{pmatrix}\\
	\end{array} $}  \\ \hline\hline

\multicolumn{4}{|c}{$\mathbf{2'} \otimes \mathbf{3'} = \mathbf{2''} \oplus \mathbf{4}$} & \multicolumn{4}{|c}{$\mathbf{2''} \otimes \mathbf{3} = \mathbf{2''} \oplus\mathbf{4}$} & \multicolumn{4}{|c|}{$\mathbf{2''} \otimes \mathbf{3'} = \mathbf{2'} \oplus \mathbf{4}$}  \\ \hline
\multicolumn{4}{|c}{  $ \begin{array}{l}
		\mathbf{2''} \sim \begin{pmatrix}
			\alpha_1 \beta_1+\sqrt{2} \alpha_2 \beta_2 \\
			\alpha_2 \beta_1-\sqrt{2} \alpha_1 \beta_3 \\
		\end{pmatrix} \\
		\mathbf{4} \sim \begin{pmatrix}
			\sqrt{2} \alpha_2 \beta_1+\alpha_1 \beta_3 \\
			\sqrt{3} \alpha_2 \beta_3 \\
			\alpha_2 \beta_2-\sqrt{2} \alpha_1 \beta_1 \\
			-\sqrt{3} \alpha_1 \beta_2 \\
		\end{pmatrix}\\
	\end{array} $ } &
\multicolumn{4}{|c}{ $ \begin{array}{l}
		\mathbf{2''} \sim \begin{pmatrix}
			\sqrt{2} \alpha_2 \beta_2-\alpha_1 \beta_1 \\
			\alpha_2 \beta_1+\sqrt{2} \alpha_1 \beta_3 \\
		\end{pmatrix}\\
		\mathbf{4} \sim \begin{pmatrix}
			\sqrt{2} \alpha_2 \beta_1-\alpha_1 \beta_3 \\
			\sqrt{3} \alpha_2 \beta_3 \\
			\sqrt{2} \alpha_1 \beta_1+\alpha_2 \beta_2 \\
			\sqrt{3} \alpha_1 \beta_2 \\
		\end{pmatrix}\\
	\end{array} $} &
\multicolumn{4}{|c|}{ $\begin{array}{l}
		\mathbf{2'} \sim \begin{pmatrix}
			\alpha_1 \beta_1-\sqrt{2} \alpha_2 \beta_2 \\
			\alpha_2 \beta_1+\sqrt{2} \alpha_1 \beta_3 \\
		\end{pmatrix} \\
		\mathbf{4} \sim \begin{pmatrix}
			\sqrt{3} \alpha_1 \beta_2 \\
			\sqrt{2} \alpha_1 \beta_1+\alpha_2 \beta_2 \\
			-\sqrt{3} \alpha_2 \beta_3 \\
			\alpha_1 \beta_3-\sqrt{2} \alpha_2 \beta_1 \\
		\end{pmatrix} \\
	\end{array} $}  \\ \hline\hline

\multicolumn{4}{|c}{$\mathbf{2} \otimes \mathbf{4} = \mathbf{2'} \oplus \mathbf{2''} \oplus \mathbf{4}$} & \multicolumn{4}{|c}{$\mathbf{2'} \otimes \mathbf{4} = \mathbf{2} \oplus \mathbf{3} \oplus \mathbf{3'}$} & \multicolumn{4}{|c|}{$\mathbf{2''} \otimes \mathbf{4} = \mathbf{2} \oplus \mathbf{3} \oplus \mathbf{3'}$}  \\ \hline
\multicolumn{4}{|c}{  $ \begin{array}{l}
	\mathbf{2'} \sim \begin{pmatrix}
		\alpha_2 \beta_3-\alpha_1 \beta_2 \\
		\alpha_2 \beta_1+\alpha_1 \beta_4 \\
	\end{pmatrix} \\
 \mathbf{2''} = \begin{pmatrix}
		\alpha_2 \beta_2+\alpha_1 \beta_3 \\
		\alpha_2 \beta_4-\alpha_1 \beta_1 \\
  \end{pmatrix} \\
   \mathbf{4} \sim \begin{pmatrix}
	\alpha_1 \beta_1+\alpha_2 \beta_4 \\
	-\alpha_1 \beta_2-\alpha_2 \beta_3 \\
	\alpha_1 \beta_3-\alpha_2 \beta_2 \\
	\alpha_2 \beta_1-\alpha_1 \beta_4 \\
 \end{pmatrix} \\
	\end{array} $ } &
\multicolumn{4}{|c}{ $ \begin{array}{l}
 \mathbf{2} \sim \begin{pmatrix}
	\alpha_2 \beta_3-\alpha_1 \beta_1 \\
	\alpha_2 \beta_2+\alpha_1 \beta_4 \\
\end{pmatrix}\\
\mathbf{3} \sim \begin{pmatrix}
	\sqrt{2} \alpha_1 \beta_1+\sqrt{2} \alpha_2 \beta_3 \\
	\alpha_1 \beta_3+\sqrt{3} \alpha_2 \beta_4 \\
	\sqrt{3} \alpha_1 \beta_2-\alpha_2 \beta_1 \\
\end{pmatrix}\\
\mathbf{3'} \sim \begin{pmatrix}
	\sqrt{2} \alpha_2 \beta_2-\sqrt{2} \alpha_1 \beta_4 \\
	\sqrt{3} \alpha_2 \beta_1+\alpha_1 \beta_2 \\
	\alpha_2 \beta_4-\sqrt{3} \alpha_1 \beta_3 \\
\end{pmatrix}\\
	\end{array} $} &
\multicolumn{4}{|c|}{ $\begin{array}{l}
\mathbf{2} \sim \begin{pmatrix}
	\alpha_1 \beta_4-\alpha_2 \beta_2 \\
	\alpha_1 \beta_1+\alpha_2 \beta_3 \\
\end{pmatrix} \\
\mathbf{3} \sim \begin{pmatrix}
	\sqrt{2} \alpha_2 \beta_2+\sqrt{2} \alpha_1 \beta_4 \\
	\sqrt{3} \alpha_2 \beta_1-\alpha_1 \beta_2 \\
	\sqrt{3} \alpha_1 \beta_3+\alpha_2 \beta_4 \\
\end{pmatrix} \\
\mathbf{3'} \sim \begin{pmatrix}
	\sqrt{2} \alpha_2 \beta_3-\sqrt{2} \alpha_1 \beta_1 \\
	\sqrt{3} \alpha_2 \beta_4-\alpha_1 \beta_3 \\
	-\alpha_2 \beta_1-\sqrt{3} \alpha_1 \beta_2 \\
\end{pmatrix} \\
	\end{array} $}  \\ \hline\hline

	\end{tabular} }
\caption{\label{tab:GL23_CG-1st}Multiplication rules of different irreducible representations of $GL(2, 3)$ and the corresponding Clebsch-Gordan coefficients.}
\end{table}

\begin{table}[ht!]
\centering
\resizebox{0.98\textwidth}{!}{
\begin{tabular}{|c|c|c|c|c|c|c|c|c|c|c|c|}\hline\hline
\multicolumn{6}{|c}{ $\mathbf{3} \otimes \mathbf{3} = \mathbf{3'} \otimes \mathbf{3'} = \mathbf{1_s} \oplus \mathbf{2_s} \oplus \mathbf{3_a} \oplus \mathbf{3'_s}$} & \multicolumn{6}{|c|}{ $\mathbf{3} \otimes \mathbf{3'} = \mathbf{1'} \oplus \mathbf{2} \oplus \mathbf{3} \oplus \mathbf{3'}$} \\ \hline
\multicolumn{6}{|c}{ $\begin{array}{l}
\mathbf{1_s} \sim \alpha_1 \beta_1+\alpha_3 \beta_2+\alpha_2 \beta_3 \\
\mathbf{2_s} \sim \begin{pmatrix}
	\alpha_3 \beta_2+\alpha_2 \beta_3-2 \alpha_1 \beta_1 \\
	\sqrt{3} \alpha_2 \beta_2+\sqrt{3} \alpha_3 \beta_3 \\
\end{pmatrix} \\
\mathbf{3_a} \sim \begin{pmatrix}
	\alpha_3 \beta_2-\alpha_2 \beta_3 \\
	\alpha_2 \beta_1-\alpha_1 \beta_2 \\
	\alpha_1 \beta_3-\alpha_3 \beta_1 \\
\end{pmatrix}  \\
\mathbf{3'_s} \sim \begin{pmatrix}
	\alpha_3 \beta_3-\alpha_2 \beta_2 \\
	\alpha_3 \beta_1+\alpha_1 \beta_3 \\
	-\alpha_2 \beta_1-\alpha_1 \beta_2 \\
\end{pmatrix} \\
\end{array} $} &
\multicolumn{6}{|c|}{ $\begin{array}{l}
\mathbf{1'} \sim \alpha_1 \beta_1+\alpha_3 \beta_2+\alpha_2 \beta_3 \\
\mathbf{2} \sim \begin{pmatrix}
	\sqrt{3} \alpha_2 \beta_2+\sqrt{3} \alpha_3 \beta_3 \\
	2 \alpha_1 \beta_1-\alpha_3 \beta_2-\alpha_2 \beta_3 \\
\end{pmatrix} \\
\mathbf{3} \sim \begin{pmatrix}
	\alpha_3 \beta_3-\alpha_2 \beta_2 \\
	\alpha_3 \beta_1+\alpha_1 \beta_3 \\
	-\alpha_2 \beta_1-\alpha_1 \beta_2 \\
\end{pmatrix}  \\
\mathbf{3'} \sim \begin{pmatrix}
	\alpha_3 \beta_2-\alpha_2 \beta_3 \\
	\alpha_2 \beta_1-\alpha_1 \beta_2 \\
	\alpha_1 \beta_3-\alpha_3 \beta_1 \\
\end{pmatrix} \\
\end{array} $}  \\ \hline \hline

\multicolumn{6}{|c}{ $\mathbf{3} \otimes \mathbf{4} = \mathbf{2'} \oplus \mathbf{2''} \oplus \mathbf{4_1} \oplus \mathbf{4_2}$} & \multicolumn{6}{|c|}{ $\mathbf{3'} \otimes \mathbf{4} = \mathbf{2'} \oplus \mathbf{2''} \oplus \mathbf{4_1} \oplus \mathbf{4_2}$} \\ \hline
\multicolumn{6}{|c}{ $\begin{array}{l}
\mathbf{2'} \sim \begin{pmatrix}
-\sqrt{3} \alpha_3 \beta_1-\sqrt{2} \alpha_1 \beta_2-\alpha_2 \beta_4 \\
\alpha_3 \beta_2-\sqrt{3} \alpha_2 \beta_3-\sqrt{2} \alpha_1 \beta_4 \\
\end{pmatrix} \\
\mathbf{2''} \sim \begin{pmatrix}
	\sqrt{2} \alpha_1 \beta_3+ \sqrt{3}\alpha_3 \beta_4-\alpha_2\beta_1 \\
	\sqrt{2} \alpha_1 \beta_1+ \sqrt{3}\alpha_2 \beta_2+\alpha_3\beta_3 \\
\end{pmatrix} \\
\mathbf{4_1} \sim \begin{pmatrix}
	\sqrt{2} \alpha_3 \beta_3-\alpha_1 \beta_1 \\
	\alpha_1 \beta_2-\sqrt{2} \alpha_2 \beta_4 \\
	\sqrt{2} \alpha_2 \beta_1+\alpha_1 \beta_3 \\
	-\sqrt{2} \alpha_3 \beta_2-\alpha_1 \beta_4 \\
\end{pmatrix} \\
\mathbf{4_2} \sim \begin{pmatrix}
	3 \alpha_1 \beta_1-\sqrt{6} \alpha_2 \beta_2 \\
	\alpha_1 \beta_2+2 \sqrt{2} \alpha_2 \beta_4-\sqrt{6} \alpha_3 \beta_1 \\
	\sqrt{6} \alpha_3 \beta_4-3 \alpha_1 \beta_3 \\
	2 \sqrt{2} \alpha_3 \beta_2+\sqrt{6} \alpha_2 \beta_3-\alpha_1 \beta_4 \\
\end{pmatrix} \\
\end{array} $} &
\multicolumn{6}{|c|}{ $\begin{array}{l}
\mathbf{2'} \sim \begin{pmatrix}
	\sqrt{2} \alpha_1 \beta_3+\sqrt{3} \alpha_3 \beta_4-\alpha_2 \beta_1\\
	-\sqrt{2} \alpha_1 \beta_1-\sqrt{3} \alpha_2 \beta_2-\alpha_3 \beta_3 \\
\end{pmatrix} \\
\mathbf{2''} \sim \begin{pmatrix}
	\sqrt{3} \alpha_3 \beta_1+\sqrt{2} \alpha_1 \beta_2+\alpha_2 \beta_4 \\
	\alpha_3 \beta_2-\sqrt{3} \alpha_2 \beta_3-\sqrt{2} \alpha_1 \beta_4 \\
\end{pmatrix} \\
\mathbf{4_1} \sim \begin{pmatrix}
	2 \sqrt{2} \alpha_3 \beta_2+\sqrt{6} \alpha_2 \beta_3-\alpha_1 \beta_4 \\
	\sqrt{6} \alpha_3 \beta_4-3 \alpha_1 \beta_3 \\
	\sqrt{6} \alpha_3 \beta_1-\alpha_1 \beta_2-2 \sqrt{2} \alpha_2 \beta_4 \\
	\sqrt{6} \alpha_2 \beta_2-3 \alpha_1 \beta_1 \\
\end{pmatrix} \\
\mathbf{4_2} \sim \begin{pmatrix}
	\sqrt{2} \alpha_3 \beta_2+\alpha_1 \beta_4 \\
	-\sqrt{2} \alpha_2 \beta_1-\alpha_1 \beta_3 \\
	\alpha_1 \beta_2-\sqrt{2} \alpha_2 \beta_4 \\
	\sqrt{2} \alpha_3 \beta_3-\alpha_1 \beta_1 \\
\end{pmatrix}\\
\end{array} $} \\ \hline\hline

\multicolumn{12}{|c|}{ $\mathbf{4} \otimes \mathbf{4} = \mathbf{1_{s}} \oplus \mathbf{1'_a} \oplus \mathbf{2_a} \oplus \mathbf{3_{1,a}} \oplus\mathbf{3_{2,s}} \oplus \mathbf{3'_{1,s}} \oplus \mathbf{3'_{2,s}}$ } \\ \hline
\multicolumn{12}{|c|}{ $\begin{array}{l}
\mathbf{1_s} \sim \alpha_2 \beta_1+\alpha_1 \beta_2+\alpha_4 \beta_3+\alpha_3 \beta_4 \\
\mathbf{1'_a} \sim \alpha_3 \beta_1-\alpha_1 \beta_3+\alpha_4 \beta_2-\alpha_2 \beta_4 \\
\mathbf{2_a} \sim \begin{pmatrix}
	\alpha_2 \beta_1-\alpha_1 \beta_2+\alpha_4 \beta_3-\alpha_3 \beta_4 \\
	\alpha_3 \beta_1-\alpha_1 \beta_3+\alpha_2 \beta_4-\alpha_4 \beta_2 \\
\end{pmatrix}\\
\mathbf{3_{1,a}} \sim \begin{pmatrix}
	\alpha_1 \beta_2-\alpha_2 \beta_1+\alpha_4 \beta_3-\alpha_3 \beta_4 \\
	\sqrt{2} \alpha_2 \beta_3-\sqrt{2} \alpha_3 \beta_2 \\
	\sqrt{2} \alpha_4 \beta_1-\sqrt{2} \alpha_1 \beta_4 \\
\end{pmatrix} \\
\mathbf{3_{2,s}} \sim \begin{pmatrix}
	\sqrt{2} \alpha_2 \beta_1+\sqrt{2} \alpha_1 \beta_2-\sqrt{2} \
	\alpha_4 \beta_3-\sqrt{2} \alpha_3 \beta_4 \\
	\sqrt{3} \alpha_4 \beta_4+\alpha_3 \beta_2+\alpha_2 \beta_3-\sqrt{3} \
	\alpha_1 \beta_1 \\
	\sqrt{3} \alpha_3 \beta_3+\alpha_4 \beta_1+\alpha_1 \beta_4-\sqrt{3} \
	\alpha_2 \beta_2 \\
\end{pmatrix} \\
\mathbf{3'_{1,s}} = \begin{pmatrix}
\alpha_3 \beta_1+\alpha_1 \beta_3+3 \alpha_4 \beta_2+3 \alpha_2 \beta_4 \\
2 \sqrt{2} \alpha_3 \beta_3-\sqrt{6} \alpha_4 \beta_1-\sqrt{6}\alpha_1 \beta_4 \\
-2 \sqrt{2} \alpha_1 \beta_1-\sqrt{6} \alpha_3 \beta_2-\sqrt{6}\alpha_2 \beta_3 \\
\end{pmatrix} \\
\mathbf{3'_{2,s}} = \begin{pmatrix}
2 \sqrt{2} \alpha_3 \beta_1+2 \sqrt{2} \alpha_1 \beta_3 \\
-3 \alpha_2 \beta_2-\alpha_3 \beta_3-\sqrt{3} \alpha_4 \beta_1-\sqrt{3}\alpha_1 \beta_4 \\
\alpha_1 \beta_1+3 \alpha_4 \beta_4-\sqrt{3} \alpha_3 \beta_2-\sqrt{3} \alpha_2 \beta_3 \\
\end{pmatrix} \\
	\end{array} $} \\ \hline \hline
\end{tabular} }
\caption{\label{tab:GL23_CG-2nd}Continuum of table~\ref{tab:GL23_CG-1st}. }
\end{table}

\end{appendix}

\clearpage

\providecommand{\href}[2]{#2}\begingroup\raggedright\endgroup

\end{document}